\documentclass[a4paper,twoside,notitlepage]{article}
\usepackage[applemac]{inputenc}
\usepackage{natbib,epsf,amssymb,bm}
\usepackage{amsmath}
\usepackage{epsfig}
\usepackage{epstopdf}
\usepackage{SectsV9}
\newcommand{\oxfordsize}{
    \setlength{\parindent}{15pt}
    \setlength{\headheight}{12pt}
    \setlength{\headsep}{16pt}
    \setlength{\topskip}{10pt}
    \setlength{\footskip}{12pt}
    \setlength{\textwidth}{118mm}
    \setlength{\textheight}{178mm}
    \setlength{\topmargin}{18mm}
    \setlength{\oddsidemargin}{21mm}
    \setlength{\evensidemargin}{21mm}
    \setlength{\parskip}{0pt plus1pt}}

\newcommand{\ti}[1]{\emph{#1}}

\newcommand{\oxfont}{\fontfamily{cmr}\fontshape{n}
  \fontseries{m}\fontsize{9}{10}\selectfont}
\newcommand{\absfont}{\fontfamily{cmr}\fontshape{n}
  \fontseries{m}\fontsize{8}{9}\selectfont}
\newcommand{\autfont}{\fontfamily{cmr}\fontshape{it}
  \fontseries{m}\fontsize{11}{12}\selectfont}
\newcommand{\titfont}{\fontfamily{cmss}\fontshape{n}
  \fontseries{bx}\fontsize{17}{18}\selectfont}

\newcommand{\fsize}[1]{
              \font\mrm=cmr10 at #1pt \font\mit=cmti10 at #1pt
              \font\mbf=cmbx10 at #1pt \oxfont}

\newcommand{\m}[1]{{\;\raise2pt\hbox{#1}}}

\newcommand{\reals}{\mbox{\rm I\kern-.20em R}}
\newcommand{\sreals}{\mbox{\small \rm I\kern-.20em R}}

\newcommand{\pin}[1]{\setcounter{page}{#1}}

 \newcommand{\bsnuevea}{\thispagestyle{empty}\oxfont
        \vspace*{-\headsep}\vspace*{-\headheight}\vspace*{-10pt}
        \begingroup\absfont \itshape
        \noindent BAYESIAN STATISTICS 9, 
        \newline J.~M.~Bernardo, M.~J.~Bayarri, J.~O.~Berger,
        A.~P.~Dawid,\newline D.~Heckerman, A.~F.~M.~Smith
        and M.~West (Eds.)\newline
        \copyright\ Oxford University Press, 2010
        \endgroup}

\newcommand{\aut}[1]{\vspace*{2mm}\centerline{{\autfont\scshape #1}}}
\newcommand{\loc}[1]{\vspace*{2pt}\centerline{{\itshape #1}}}
\newcommand{\email}[1]{\vspace*{2pt}\centerline{{\ttfamily #1}}}

\newcommand{\running}[2]{\markboth{\hfill {{\itshape #1}}}
 {{\itshape #2}}\hfill}

\newcommand{\info}[1]{
 \renewcommand{\thefootnote}{\fnsymbol{footnote}}
 \footnotetext[0]{\kern-1.6\parindent #1}
 \renewcommand{\thefootnote}{\arabic{footnote}}}

\renewcommand{\appendix}{\section*{Appendix}}

\newcommand{\blef}{\begin{flushleft}}
\newcommand{\elef}{\end{flushleft}}

\newcommand{\bcen}{\begin{center}}
\newcommand{\ecen}{\end{center}}

\newenvironment{tit}{\bcen\titfont}{\ecen}
\newcommand{\btit}{\vspace*{18mm}\begin{tit}}
\newcommand{\etit}{\end{tit}}

\newenvironment{cit}
{\begin{quote}\absfont}{\end{quote}}
\newcommand{\bcit}{\begin{cit}}
\newcommand{\ecit}{\end{cit}}

\newenvironment{abs}
{\begin{quote}\absfont \bcen{\scshape Summary}\ecen \vspace*{1mm} }
{\end{quote}}
\newcommand{\babs}{\begin{abs}}
\newcommand{\eabs}{\end{abs}}

\newenvironment{key}
{\begin{quote}\begin{flushleft}\absfont \ti{Keywords and Phrases:} 
  \scshape}{\end{flushleft}\end{quote}}
\newcommand{\bkey}{\begin{key}}
\newcommand{\ekey}{\end{key}\vspace*{1mm}}

\newcommand{\beqn}{\begin{eqnarray*}}
\newcommand{\eeqn}{\end{eqnarray*}}
\newcommand{\beqnn}{\begin{eqnarray}}
\newcommand{\eeqnn}{\end{eqnarray}}

\newenvironment{mat}{\left(\begin{array}}{\end{array}\right)}
\newcommand{\bmat}{\begin{mat}}
\newcommand{\emat}{\end{mat}}

\newenvironment{arr}{\begin{array}}{\end{array}}
\newcommand{\barr}{\begin{arr}}
\newcommand{\earr}{\end{arr}}

\newtheorem{teo}{Theorem}
\newcommand{\bteo}{\begin{teo}\itshape}
\newcommand{\bteon}[1]{\begin{teo}{\fsize{9}\mbf (#1).}\itshape}
\newcommand{\eteo}{\end{teo}}

\newtheorem{cor}{Corollary}[teo]
\newcommand{\bcor}{\begin{cor}\itshape}
\newcommand{\ecor}{\end{cor}}

\newtheorem{dfn}{Definition}
\newcommand{\bdfn}{\begin{dfn}\itshape}
\newcommand{\bdfnn}[1]{\begin{dfn}{\fsize{9}\mbf (#1).}\itshape}
\newcommand{\edfn}{\end{dfn}}
\newcommand{\edfnn}{\end{dfn}}

\newtheorem{exa}{{\absfont\bfseries Example}}
\newcommand{\bexa}{\begin{exa}\absfont}
\newcommand{\bexan}[1]{\begin{exa}{\fsize{8}\mbf (#1).}\absfont}
\newcommand{\eexa}{\end{exa}}

\newcommand{\bite}{\begin{itemize}}
\newcommand{\eite}{\end{itemize}}

\newcommand{\bnum}{\begin{enumerate}}
\newcommand{\enum}{\end{enumerate}}

\newenvironment{parag}{\par}{\par}
\newenvironment{dif}
  {\begin{parag}\absfont  \begin{parag}}
  {\end{parag}\end{parag}}
\newcommand{\bdif}{\begin{dif}}
\newcommand{\edif}{\end{dif}}

\newcommand{\ok}{\hfill\raise-3pt\hbox{$\square$}\vspace*{1mm}}
\newenvironment{proof}
      {\begin{dif} \noindent{\em Proof.~}}
      {\ok\vspace*{15pt}\end{dif}}
\newcommand{\bpro}{\begin{proof}}
\newcommand{\epro}{\end{proof}}

\newcommand{\capfig}[1]{
    \begin{quote}\caption{\ignorespaces\fsize{8}\mit #1}\end{quote}}

\newcommand{\btab}[1]{\vspace*{-3mm}\begin{table}[h]
      \begin{center}\absfont\capfig{{#1}}}
\newcommand{\etab}{\hline\end{tabular}
      \vspace*{-2mm}\end{center}\end{table}\par}
\newcommand{\format}[1]{\begin{tabular}{#1}\hline}
\newcommand{\btabb}[2]{\vspace*{-#2mm}\begin{table}[h]
      \begin{center}\absfont\capfig{{#1}}}
\newcommand{\etabb}[1]{\hline\end{tabular}
      \vspace*{-#1mm}\end{center}\end{table}\par}

\newcommand\pasteps[2]{\par
     \centerline{\epsfxsize=#2mm\epsfbox{./#1}}\par}

\newcommand{\bfig}[2]{\par\begin{figure}[h]
   \pasteps{#1}{#2}}
\newcommand{\efig}[1]{\capfig{#1}\vspace*{-8mm}\end{figure}}
\newcommand{\efigg}[2]{\capfig{#1}\vspace*{-#2mm}\end{figure}}

\newcommand{\bref}{\section*{References}
    \begingroup\blef\frenchspacing\absfont}
\newcommand{\eref}{\elef\endgroup}

\newcommand{\gb}{\pagebreak[3]\vspace{0pt plus 2ex}}
\newcommand{\rr}{\par\gb\vspace{1pt plus 1pt
    minus0pt}\hangindent=1.5em\hangafter=1\noindent}

\def\as#1{{\it Ann.\ Sta\-tist.~}{\bf#1}}

\def\ams#1{{\it Ann.\ Math.\ Sta\-tist.~}{\bf#1}}

\def\jasa#1{{\it J.~Amer.\ Sta\-tist.\ Assoc.~}{\bf#1}}

\def\jrssb#1{{\it J.~Roy.\ Sta\-tist.\ Soc.~B~}{\bf#1}}

\def\sc#1{{\it Sta\-tist.\ Computing~}{\bf#1}}

\def\stsc#1{{\it Sta\-tist.\ Science~}{\bf#1}}

\def\academic{New York: Academic Press}

\def\california{Berkeley: Univ. California Press}

\def\nova{New York: Nova Science}

\def\springer{Berlin: Springer}
\def\springerny{New York: Springer}

\def\val#1{\ifcase#1
 \or {\it Bayes\-ian Statistics}
  (J.~M. Ber\-nar\-do, M.~H. DeGroot, D.~V. Lind\-ley and
   A.~F.~M. Smith, eds.) Va\-len\-cia: University Press%
  \or {\it Bayes\-ian Statistics~2}
  (J.~M. Ber\-nar\-do, M.~H. DeGroot, D.~V. Lind\-ley and
   A.~F.~M. Smith, eds.), Ams\-ter\-dam: North-Holland%
  \or {\it Bayes\-ian Statistics~3}
  (J.~M. Ber\-nar\-do, M.~H. DeGroot, D.~V. Lind\-ley and
   A.~F.~M. Smith, eds.) Oxford: University Press%
  \or {\it Bayes\-ian Statistics~4}
  (J.~M. Ber\-nar\-do, J.~O. Ber\-ger, A.~P. Dawid and
   A.~F.~M. Smith, eds.) Oxford: University Press%
  \or {\it Bayes\-ian Statistics~5}
  (J.~M. Ber\-nar\-do, J.~O. Ber\-ger, A.~P. Dawid and
   A.~F.~M. Smith, eds.) Oxford: University Press%
  \or {\it Bayesian Statistics~6}
  (J.~M. Ber\-nar\-do, J.~O. Ber\-ger, A.~P. Dawid and
   A.~F.~M. Smith, eds.) Oxford: University Press%
 \or {\it Bayesian Statistics~7}
  (J.~M. Ber\-nar\-do, M.~J.~Ba\-yarri, J.~O. Ber\-ger, A.~P. Dawid, D.
Hecker\-man,
   A.~F.~M. Smith and M. West, eds.) Oxford: University Press\fi}

\def\berk#1#2{\ifcase#1
 \or {\it Proc.\ First Berkeley Symp.}
  (J.~Neyman ed.) \california%
 \or {\it Proc.\ Second Berkeley Symp.}
  (J.~Neyman ed.) \california%
 \or {\it Proc.\ Third Berkeley Symp.}~{\bf#2}
  (J.~Neyman  and E.~L.~Scott, eds.) \california%
 \or {\it Proc.\ Fourth Berkeley Symp.}~{\bf#2}
  (J.~Neyman  and E.~L.~Scott, eds.) \california%
 \or {\it Proc.\ Fifth Berkeley Symp.}~{\bf#2}
  (J.~Neyman  and E.~L.~Scott, eds.) \california%
 \or {\it Proc.\ Sixth Berkeley Symp.}~{\bf#2}
  (L. Le Cam, J.~Neyman  and E.~L.~Scott, eds.) \california\fi}

\def\applied#1{\ifcase#1
 \or {\it Applied Statistical Science I}
  (M.~Ahsanullah and  D.~S~Bhoj, eds.) \nova%
 \or {\it Applied Statistical Science II}
  (M.~Ahsanullah, ed.) \nova%
 \or {\it Applied Statistical Science III}
  (S.~E. Ahmed, M. Ahsanullah and B. K. Sinha, eds.) \nova%
 \or {\it Applied Statistical Science IV}
  (M.~Ahsanullah and F. Yildirim, eds.) \nova%
 \or {\it Applied Statistical Science V}
  (M.~Ahsanullah, J.~Kennyon and  S.~K.~Sarkar, eds.) \nova\fi}

\def\purdue#1#2{\ifcase#1\or
  {\it Statistical Decision Theory and Related Topics}
   (S.~S.~Gupta and J.~Yackel, eds.) \academic%
  \or {\it Statistical Decision Theory and Related Topics II}
   (S.~S.~Gupta and D. S.~Moore, eds.) \academic%
  \or {\it Statistical Decision Theory and Related Topics III}~{\bf#2}
  (S.~S.~Gupta and J.~O.~Berger, eds.) \academic%
  \or {\it Statistical Decision Theory and Related Topics IV}~{\bf#2}
  (S.~S.~Gupta and J.~O.~Berger, eds.) \springer%
  \or {\it Statistical Decision Theory and Related Topics V}~{\bf#2}
  (S.~S.~Gupta and J.~O.~Berger, eds.) \springer\fi}

\def\casebayes#1{\ifcase#1\or
  {\it Case Studies in Bayesian Statistics}
   (C.~Gatsonis, J.~S.~Hodges, R.~E.~Kass and N.~D.~Singpurwalla, eds.)
\springerny%
  \or {\it Case Studies in Bayesian Statistics II}
   (C.~Gatsonis, J.~S.~Hodges, R.~E.~Kass and N.~D.~Singpurwalla, eds.)
\springerny%
  \or {\it Case Studies in Bayesian Statistics III}
  (C.~Gatsonis, J.~S.~Hodges, R.~E.~Kass, R.~E.~McCulloch, P.~Rossi
    and N.~D.~Singpurwalla, eds.) \springerny%
  \or {\it Case Studies in Bayesian Statistics IV}
  (C.~Gatsonis, R.~E.~Kass, B.~Carlin, A.~Carriquiry, A.~Gelman, 
   I.~Verdinelli and M.~West, eds.) \springerny%
  \or {\it Case Studies in Bayesian Statistics V}
  (C.~Gatsonis, R.~E.~Kass, B.~Carlin, A.~Carriquiry, A.~Gelman, 
   I.~Verdinelli and M.~West, eds.) \springerny%
  \or {\it Case Studies in Bayesian Statistics VI}
  (C.~Gatsonis, R.~E.~Kass, A.~Carriquiry,  A.~Gelman, D.~Higdon, D.~K.~Pauler 
   and I.~Verdinelli, eds.) \springerny\fi}

\fsize{10}\oxfordsize\oxfont\pin{1}

\begin{document}
\bsnuevea

\newcommand{\zsqm}{Z^2_m}
\newcommand{\dt}{\delta t}
\newcommand{\eps}{\epsilon}
\newcommand{\afreq}{\omega}
\newcommand{\freq}{f}
\newcommand{\fmax}{\freq_{\rm max}}
\newcommand{\fdot}{\dot{\freq}}
\newcommand{\fdotmax}{\dot \freq_{\rm max}}
\newcommand{\per}{\tau}
\newcommand{\spgram}{{\cal P}}
\newcommand{\pgram}{periodogram}
\newcommand{\pval}{$p$-value}
\newcommand{\pvals}{$p$-values}
\newcommand{\Fermi}{{\em Fermi}}
\newcommand{\efold}{$\chi^2$-EF}
\newcommand{\like}{{\mathcal L}}
\newcommand{\spars}{{\mathcal S}}
\newcommand{\fvec}{{\bm f}}

\newcommand{\Info}{{\mathcal I}}
\newcommand{\expect}{\mathbb{E}}
\newcommand{\einfo}{\expect\Info}
\newcommand{\ent}{{\mathcal H}}

\newcommand{\ceqn}[1]{equation~(\ref{#1})}
\newcommand{\ceq}[1]{(\ref{#1})}

\newcommand\enote[1]{{$\bullet\bullet\bullet$}{\sl [#1]}{$\bullet\bullet\bullet$}}

\btit
Rotating Stars and Revolving Planets:\\
Bayesian Exploration of the Pulsating Sky
\etit

\running{T.~J.~Loredo}{The Pulsating Sky}

\aut{Thomas~J.~Loredo}
\loc{Cornell University, USA}
\email{loredo@astro.cornell.edu}

\info{Work reported here was funded in part by NASA grants NAG 5-1758 and
NNX09AK60G, by NASA's {\em Space Interferometry Mission}, and by NSF grants
AST-0507254 and AST-0507589.}

\babs
I describe ongoing work on development of Bayesian methods for exploring
periodically varying phenomena in astronomy, addressing two classes of
sources: pulsars, and extrasolar planets (exoplanets).  For pulsars, the
methods aim to detect and measure periodically varying signals in data
consisting of photon arrival times, modeled as non-homogeneous Poisson point
processes.  For exoplanets, the methods address detection and estimation of
planetary orbits using observations of the reflex motion ``wobble'' of a host
star, including adaptive scheduling of observations to optimize inferences.
\eabs

\bkey
Time series; Poisson point processes; Harmonic analysis;
Periodograms; Experimental design; Astronomy; Pulsars; Extrasolar planets
\ekey

\section{Introduction}

In his famous sonnet, ``Bright Star'' (1819), John Keats addresses a
star, lamenting of the transience of human emotions---and of human life
itself---in contrast to the star's immutability:


\begin{quote}
Bright star, would I were steadfast as thou art---\\
Not in lone splendor hung aloft the night\\
And watching, with eternal lids apart,\\
Like nature's patient, sleepless Eremite\ldots\\
\ldots yet still steadfast, still unchangeable\ldots
\end{quote}

\noindent Many decades later, Robert Frost alluded to ``Keats' Eremite'' in
his poem, ``Choose Something Like a Star'' (1947).  The poet queries a
star (``the fairest one in sight''), pleading for a celestial lesson
that ``we can learn/By heart and when alone repeat.''  He finds,

\begin{quote}
It gives us strangely little aid,\\
But does tell something in the end\ldots\\
It asks of us a certain height,\\
So when at times the mob is swayed\\
To carry praise or blame too far,\\
We may choose something like a star\\
To stay our minds on and be staid. 
\end{quote}

Both poets invoke a millenia-long, cross-cultural tradition of finding in the
``fixed stars'' a symbol of constancy; sometimes cold, sometimes comforting.
But these poems of Keats and Frost bookmark a period of enormous change in our
understanding of cosmic variability.

Already by Keats's time---marked by the discovery of invisible infrared and
ultraviolet light in the Sun's spectrum (by Herschel, 1800, and Ritter, 1801),
and by the dawn of stellar spectroscopy (Fraunhofer, 1823)---astronomers were
discovering that there was quite literally ``more than meets the eye'' in
starlight.  Later in the 19th century, long-exposure astrophotography extended
the reach of telescopes and spectroscopes to ever dimmer and farther objects,
and provided reproducibly precise records that enabled tracking of properties
over time.  In the 20th century, advances in optics and new detector
technologies extended astronomers' ``vision'' to wavelengths and frequencies
much further beyond the narrow range accessible to the retina.  By
mid-century, some of these tools became capable of short-time-scale
measurements.  Simultaneous with these technological developments were
theoretical insights, most importantly from nuclear physics, that unveiled the
processes powering stars, processes with finite lifetimes, predicting stellar
evolution and death, including the formation of compact, dense stellar
remnants.

By the time of Frost's death (1963), astronomers had come to see stars as
ever-changing things, not only on the inhumanly long billion-year time scales
of stellar evolution, but even over humanly accessible periods of years,
months, and days.  Within just a decade of Frost's death, the discoveries of
pulsars, X-ray transients, and gamma-ray bursts revealed that solar-mass-scale
objects were capable of pulsing or flashing on timescales as small as {\em
milliseconds}.

We now know that the ``fixed'' stars visible to the naked eye represent a
highly biased cross-sectional sample of an evolving population of great
heterogeneity.  The more complete astronomical census made possible by modern
astronomical instrumentation reveals the heavens to be as much a place of
dramatic---sometimes violent---change as a harbor of steady luminance.
The same instrumentation also reveals subtle but significant change even
among some of the visible ``fixed'' stars.

Here I will point a Bayesian statistical telescope of sorts at one particular
area of modern time-domain astronomy:  periodic variability.  Even this small
area encompasses a huge range of phenomena, as is the case in other
disciplines studying periodic time series.  I will focus on two small but
prominent corners of periodic astronomy:  studies of {\em pulsars} (rapidly
rotating neutron stars) and of {\em extrasolar planets} (``exoplanets,''
planetary bodies revolving around other suns).  New and upcoming
instrumentation are producing rich data sets and challenging statistical
inference problems in both pulsar and exoplanet astronomy.  Bayesian methods
are well-suited to maximizing the science extracted from the exciting new
data.

The best-known and most influential statistical methods for detecting and
characterizing periodic signals in astronomy use {\em periodograms}.  In the
next section I will take a brief, Bayesian look at periodograms; they shed
light on important issues common to many periodic time series problems, such
as strong multimodality in likelihood functions and posterior densities.  In
\S~3 I describe detection and measurement of pulsars using data that report
precise arrival times of individual light quanta (photons), including 
Bayesian approaches to arrival time series analysis using parametric and
semiparametric inhomogeneous Poisson point process models.  In \S~4 I turn to
exoplanets, where the most productive detection methods as of this writing
find planets that are too dim to see directly by looking for the reflex motion
``wobble'' of their host stars. Here the data are irregularly sampled time
series with additive noise, with very accurate but highly nonlinear parametric
models for the underlying orbital motion.  I will briefly describe some key
inference problems (e.g., planet detection and orbit fitting), but I will
focus on application of Bayesian experimental design to the problem of
adaptive scheduling of the costly observations of these systems.  A running
theme is devising Bayesian counterparts to well-known frequentist methods, and
then using the Bayesian framework to add new capability not so readily
achieved with conventional approaches.  A final section offers some closing
perspectives.

\section{Periodograms and multimodality}

Suppose we have data consisting of samples of a time-dependent signal, $f(t)$,
corrupted by additve noise; suppose the sample times, $t_i$ ($i=1$ to $N$) are
uniformly spaced in time, with spacing $\dt$ and total duration $T = t_N -
t_1$.  The measured data, $d_i$, are related to the signal by,
\beqnn
d_i = f(t_i) + \eps_i,
\label{d-f-e}
\eeqnn
where $\eps_i$ denotes the unknown noise contribution to sample $i$.  If
we suspect the signal is periodic with period $\per$ and frequency
$\freq = 1/\per$, a standard statistical tool for
assessing periodic hypotheses is the {\em Schuster periodogram}
(Schuster 1898), a continuous function of the unknown angular
frequency of the signal, $\afreq = 2\pi\freq$:
\beqnn
\spgram(\afreq) = \frac{1}{N} \left[ C^2(\afreq) + S^2(\afreq) \right],
\label{spgram-def}
\eeqnn
where $C$ and $S$ are projections of the data onto cosine and sine functions;
\beqnn
C(\afreq) = \sum_i d_i \cos(\afreq t_i),\qquad
S(\afreq) = \sum_i d_i \sin(\afreq t_i).
\label{CS-def}
\eeqnn
Using trigonometric identities one can show that
\beqnn
\spgram(\afreq)
  = \frac{1}{N} \left| \sum_i d_i e^{i\afreq t_i} \right|^2.
\label{pgram-dft}
\eeqnn
Thus the \pgram\ is the squared magnitude of a quantity like the
discrete Fourier transform (DFT), but considered as a continuous function of
frequency; accordingly, the \pgram\ ordinate is often called the {\em power}
at frequency $\afreq$.  The \pgram\ is a periodic function of $\afreq$, with
period $2\pi/\dt$, and it is reflection-symmetric about the midpoint of each
such frequency interval; these symmetries reflect the aliasing of signals with
periods smaller than twice the interval between samples (i.e., periods for
which the data are sampled below the Nyquist rate).  We will assume the
angular frequencies of interest have $\afreq \in (0, \pi/\dt)$; equivalently,
$\freq \in (0, 1/2\dt)$.


Suppose the available information justifies assigning independent standard
normal probability densities for the $\eps_i$.  Then the \pgram\ has several
simple and useful properties.  Under the null hypothesis ($H_0$) of a constant
signal, $f(t) = 0$, the {\em Fourier frequencies}, $\freq_j = j/T$ ($j=1$ to
$N/2$), play a special role.  The $N_F = N/2$ values
$\{\spgram(2\pi\freq_j)\}$ are statistically independent; the probability
distribution for each value of $2\spgram(2\pi\freq_j)$ is $\chi^2_2$ (i.e.,
the periodogram values themselves have exponential distributions).  The
independence implies that the continuous function $\spgram(\afreq)$ may be
expected to have significant structure on angular frequency scales $\sim
2\pi/T$ (or $1/T$ in $\freq$), the Fourier spacing.


The best-known use of periodograms in astronomy is for nonparametric periodic
signal detection via a significance test that attempts to reject the
null.  The simplest procedure examines $\spgram(\afreq_j)$ at the Fourier
frequencies to find the highest power.  From the $\chi^2_2$ null distribution
a \pval\ may be calculated, say, $p_1$.  The overall \pval, $p$, must account for
examination of $N/2$ independent \pgram\ ordinates; a Bonferroni correction
leads to $p \approx N_F p_1$ (for small $p_1$).  When $p$ is small (say,
$p<0.01$), one claims there is significant evidence for a periodic signal;
astronomers refer to $p$ as the {\em significance level} associated with
the claimed detection.

In practice, when a periodic signal is present, its frequency will not
correspond to a Fourier frequency, reducing power (in the Neyman-Pearson
sense).  Thus one {\em oversamples} by a factor $M$, examining the \pgram\ at
$M\times N_F$ frequencies with a sub-Fourier frequency spacing, $\delta\afreq
= 1/(MT)$ with $M$ typically a small integer.  The multiple testing correction is now
more complicated because the \pgram\ ordinates are no longer independent
random variables; an appropriate factor may be found via Monte Carlo
simulation, though simple rules-of-thumb are often used.


There is a complementary {\em parametric} view of the periodogram, arising
from time-domain harmonic modeling of the signal.  As a simple periodic model
for the signal, consider a sinusoid of unknown frequency, phase $\phi$, and
amplitude, $A$: $f(t) = A \cos(\afreq t - \phi)$.  Least squares (LS) fitting
of this single harmonic to the data examines the sum of squared residuals,
\beqnn
Q(\afreq, A, \phi) = \sum_i \left[d_i - A\cos(\afreq
t_i-\phi)\right]^2.
\label{Q-def}
\eeqnn
The log-likelihood function, using the standard normal noise model, is
$L(\afreq, A, \phi) = - \frac{1}{2}Q(\afreq, A, \phi)$, so the same sum plays
a key role in maximum likelihood (ML) fitting.  For a given candidate
frequency, we can analytically calculate the conditional (on $\afreq$) LS
estimates of the amplitude and phase, $\hat A(\afreq)$ and $\hat\phi(\afreq)$.
 To estimate the frequency, we can examine the profile statistic, $Q_p(\afreq)
= Q(\afreq, \hat A(\afreq), \hat\phi(\afreq))$; the best-fit frequency
minimizes this (i.e., maximizes the profile likelihood). The profile statistic
is closely connected to the \pgram; one can show
\beqnn
Q_p(\afreq) = \hbox{Const.} - \spgram(\afreq),
\label{LS-pgram}
\eeqnn
where the constant is a function of the data but not the parameters.
A corollary of this intimate
connection between parametric harmonic analysis and periodograms is that
the strong variability of the (nonparametric) periodogram implies strong multimodality
of the harmonic model likelihood function (and hence of the posterior distribution
in Bayesian harmonic analysis), on frequency scales $\sim 1/T$.

In astronomy it is frequently the case that phenomena are not sampled
uniformly, if only due to the constraint of night-sky observation
and the vagaries of telescope scheduling and weather.
The \pgram/least squares connection provided the key to generalizing
\pgram-based nonparametric periodic signal detection to
{\em non}uniformly sampled data.  Lomb (1976) and Scargle (1982) took
the connection as a defining property of the periodogram, leading
to a natural generalization for nonuniform data called the
{\em Lomb-Scargle periodogram} (LSP).  Though developed for analysis
of astronomical data, the LSP is now a widely used tool in time
series analysis across many disciplines.

Only recently was the Bayesian counterpart to this worked out, by
Jaynes (1987) and Bretthorst (1988).  Instead of {\em maximizing} a likelihood
function over amplitude and phase, they ``do the Bayesian thing'' and
{\em marginalize} over these parameters.  The logarithm of the marginal
density for the frequency is then proportional to the periodogram; for
irregularly sampled data, there is a similar connection to the LSP
(Bretthorst 2001).  But this was more than a rediscovery of earlier results in
new clothing.  From within a Bayesian framework, the calculations for converting
\pgram\ values into probability statements about the signal differ starkly from
their frequentist spectral analysis counterparts.

The most stark difference appears, not in parameter estimation, but in signal
detection via model comparison.  The conditional odds for a periodic signal
being present at an a priori known frequency is approximately an exponential
of the periodogram.  But the frequency is never known precisely a priori.  For
detecting new periodic sources, one must perform a ``blind search'' over a
large frequency range.  Even for recovering a known signal in new data, the
(predictive) frequency uncertainty, based on earlier measurements, is
typically considerable. In Bayesian calculations, frequency uncertainty is
accounted for by calculating {\em marginal} rather than {\em maximum}
likelihoods, with the averaging over frequency in the marginalization integral
being the counterpart to Bonferroni correction.  There is no special role for
Fourier frequencies in this calculation, either in location or in number; in
fact, one wants to evaluate the periodogram at as many frequencies as needed
to accurately calculate the integral under the {\em continuous} periodgram
(exponentiated).  Oversampling, to get an accurate integral, adds no new
complication to the calculation.

A further difference comes from quantifying evidence for a signal with
the probability for a periodic hypothesis, instead of a \pval\ quantifying
compatibility with the null.  Very commonly, astronomers observe {\em
populations} of sources; detection and measurement of individual sources is
merely a stepping stone toward characterization of the population as a whole.
Signal probabilities (or marginal likelihoods and Bayes factors) facilitate
population modeling via multilevel (hieararchical) models.  Roughly speaking,
marginal likelihoods provide a weighting that allows one to account for
detection uncertainty in population inferences; e.g., when inferring the
number of dim sources, a large number of marginal detections may provide
strong evidence for a modest number of sources, even though one may not be
able to specify precisely which of the candidate sources are actual sources.
In contrast, population-level inference is awkward and challenging when \pvals\ are
used to quantify the evidence for a signal.  For example, one might attempt to
use false-discovery rate control to find a threshold \pval\ corresponding to a
desired limit on the number of false claimed detections within a population
(see Hopkins et al.\ 2002 for an astronomical example).  But the (unknown) actual false
detections will be preferentially clustered at low signal levels, corrupting
population-level inferences of the distribution of signal amplitudes.

A valid criticism of the Bayesian approach is the need to employ an explicit
signal model, here a single sinusoid, raising concern about behavior for
signals not resembling the model.  A frequentist nonparametric ``omnibus''
test that focuses on rejection of a null appears more robust.  But recent
theoretical insights into the capabilities of frequentist hypothesis tests
ameliorate this criticism.

Imagine an omnibus goodness-of-fit test that aims to detect periodicity by
testing for arbitrary (periodic) departures from a constant signal. Set the
test size $\alpha$ (the maximum \pval\ we will accept as indicating the
actual signal is not constant) to be small, $\alpha \ll 1$, corresponding to a
small expected ``false alarm'' rate for a Neyman-Pearson test (it is 
worth emphasizing that the observed \pval\ itself is {\em not} a false alarm
rate, despite increasingly frequent use of such terminology in the astronomy
literature).  We would like the test power $\beta$ (the long-run rate of
rejection of the null when a non-constant signal is present) to be as near unity
as possible for arbitrary non-constant signals. Janssen (2000) and Lehman and
Romano (2005; LR05) examine the power of such omnibus tests over all local
alternatives (i.e., alternatives, described in terms of a basis, in a region
of hypothesis space about the null shrinking in size like $1/\sqrt{N}$ for
data sets of size $N$).  They show that $\beta\approx\alpha$ for all
alternatives except for those along a finite number of directions in
hypothesis space (independent of $N$).  As a result, ``A proper choice of test
must be based on some knowledge of the possible set of alternatives for a
given experiment'' (LR05).  Freedman (2009) proves a complimentary theorem
showing that, for any choice of test, there are some {\em remote} alternatives
(i.e., not in a shrinking neighborhood of the null) for which $\beta \approx
0$.  As a consequence of these and related results, he concluded,
``Diagnostics cannot have much power against general alternatives.''

These results are changing practice in construction of frequentist tests.
Instead of devising clever statistics that embody an intuitively appealing
``generic'' measure of non-uniformity, statisticians are turning to the
practice of specifying an explicit family of alternatives (e.g., via a
specific choice of basis), and deriving tests that concentrate power within
the chosen family (e.g., Bickel et al.\ 2006).  An example in astronomy is the
work of Bickel, Kleijn and Rice (2008) on pulsar detection, using a Fourier
basis.  These developments indicate that, one way or another, one had better
consider specific alternatives to the null.  In this respect, parametric
Bayesian model comparison (with a prior over a broad parametric family) and
nonparametric frequentist testing do not seem very far apart.  With this
perspective, we can see the links between the \pgram\ and both frequentist and
Bayesian harmonic analysis as exposing the choice of alternatives implicit in
\pgram-based periodic signal detection.

Summarizing, some key points from this brief look at periodograms, which will
guide subsequent developments, are:  (1)~We expect the likelihood (and thus
the posterior) will be highly multimodal in the frequency dimension. (2)~The
scale of variability of the likelihood in the frequency dimension will be
$\sim 1/T$.  For problems with long-duration datasets and significant prior
frequency uncertainty, exploring the frequency dimension will be challenging.
(3)~A key difference between Bayesian and frequentist approaches arises from
how frequency uncertainty (and other parameter uncertainty) is handled, e.g.,
whether one maximizes and then corrects for multiple tests, or marginalizes,
letting probability averaging implicitly account for the parameter space size.


\section{Pulsar Science With Sparse Arrival Time Series}

\begin{quote}\setlength{\leftskip}{3em}
So near you are, summer stars,\\
So near, strumming, strumming,\\
So lazy and hum-strumming.\\
\nobreak
\hbox{\hskip5em} ---{\em Carl Sandberg}
\end{quote}

In 1967, Jocelyn Bell, a graduate student of the radio astronomer Anthony
Hewish, was monitoring radio observations of the sky that combined good
sensitivity with fast (sub-second) time resolution.  She made a startling
discovery:  a celestial source was emitting a strong periodic signal with a
period of {\em less than a second}.  It is hard to appreciate today just how
shocking this discovery was.  Theoretical astrophysicist Philip Morrison
recalled the early reaction to the news in an interview for the American
Institute of Physics:\footnote{Excerpt from the AIP ``Moments of Discovery''
web exhibit at
{\tt http://www.aip.org/history/mod/pulsar/pulsar1/01.html}.}

\begin{quote}
I remember myself meeting at the airport a friend who just returned from Great
Britain, an astronomer.  And he said, ``Have you heard the latest? \ldots
They've got something that pulses every
second---a stellar signal that pulses every second.'' I said, ``Oh, that couldn't
be true!'' ``Yes,'' he said, ``it's absolutely true. They announced it recently.
They've studied it for about five or six months. It's extraordinary.''

\ldots[T]hey sat on these results for several months, because the
whole thing was so extraordinary and so unexpected, that they didn't
want to release it until they had a chance to confirm it.

The reason of course is quite simple. We think of the stars quite
sensibly as being---well we say the fixed stars---as being eternal,
long-lived, everlasting.  And even though we know that's not 100\%
true---that the star sometimes explodes a little bit, making a nova, or
explodes disruptively flinging itself apart entirely, making a
supernova---still those are not really fast events from a human time
scale.  If they take a few seconds or a day, that would be remarkable for
a star.  You don't see much happening on the stars in a second\ldots.

[W]e knew something
remarkable was going on and people gave it a name, pulsar\ldots\ 
of course the whole astronomical community was galvanized in looking at it.
\end{quote}

We now understand pulsars to be rapidly rotating, highly magnetized neutron
stars, dense remnants of the cores of massive stars, with masses somewhat
larger than that of the Sun, but occupying a nearly spherical volume only
$\sim 10$~km in radius, and hence with a density similar to that of an atomic
nucleus.  The pulsations are due to radiative processes near the star that get
their energy from the whirling magnetic field, which acts like a generator,
accelerating charged particles to high energies.  The particles radiate in 
beams rotating with the star; the observed pulsars are those whose beams sweep
across the line of sight to Earth, in the manner of a lighthouse.  The fastest
pulsars rotate about 700 times a second; more typical pulsars have periods of
order a second.  If we could hear the variation in intensity of the light they
emit, the slower ones would sound like a ticking clock (of extraordinary
accuracy); the faster ones would hum and whine.

To date about two thousand pulsars have been discovered; ongoing surveys
continue to add to the number. The majority of pulsars pulse in radio waves. 
But a number of them also pulse in higher energy radiation:  visible light,
X~rays, and gamma~rays. Recently, a small number of {\em radio-quiet pulsars}
have been found that pulse only in high energy radiation.  Figure~1 shows
folded light curves---radiation intensity vs.\ time, with time measured in
fractions of the period---for several pulsars observed across the
electromagnetic spectrum.  There are clear differences in the light curves for
a particular pulsar across energy ranges, indicating that different physical
processes, probably in spatially distinct regions, produce the various types
of emission.  Astronomers are trying to detect and measure as many pulsars as
possible, across the electromagnetic spectrum, to characterize pulsar emission
as a population phenomenon, pooling information from individual sources to
unravel the physics and geometry of pulsar emission and how it may relate to
the manner of stellar death and the magnetic and material environments of
stars.

\bfig{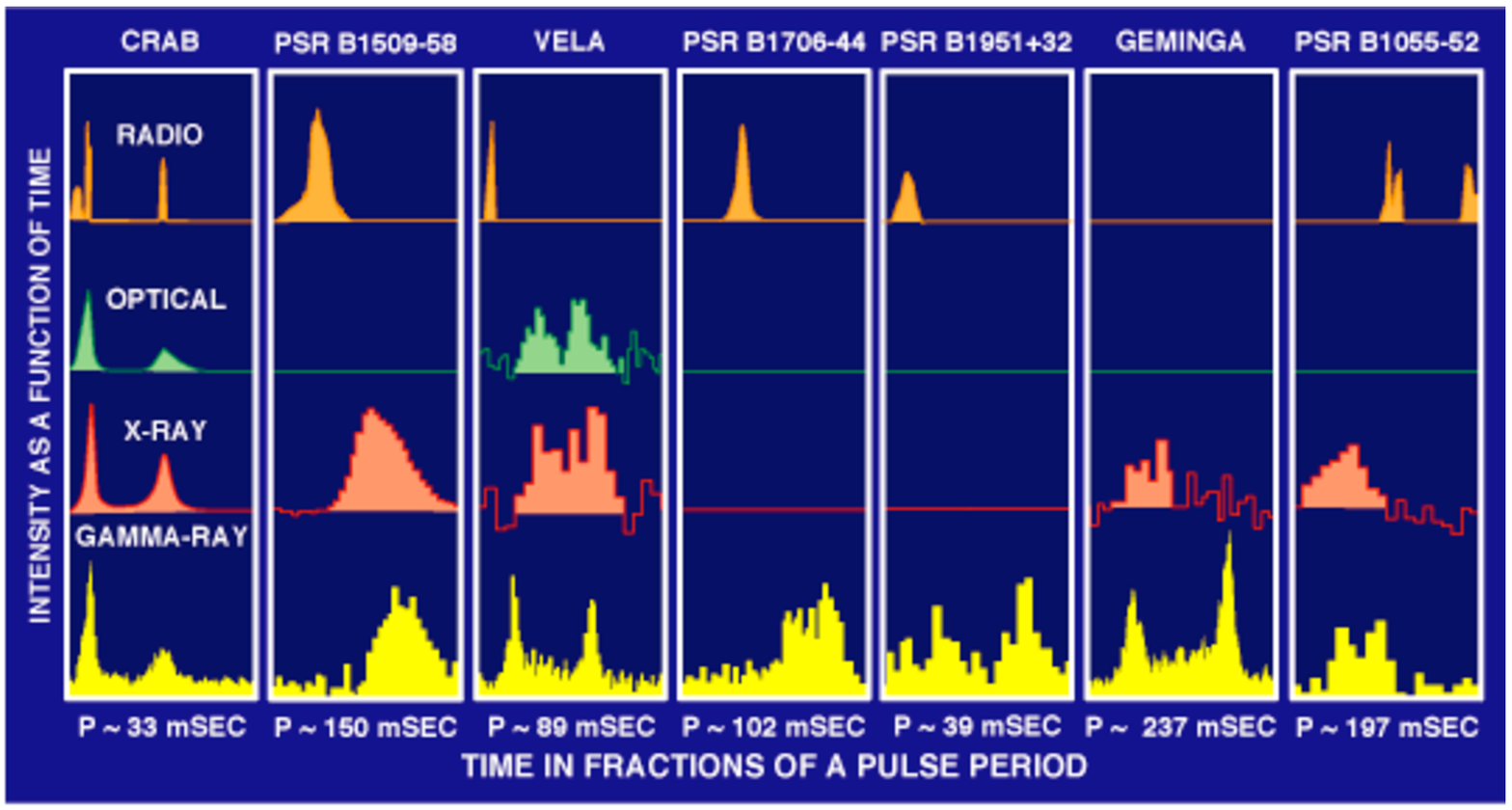}{100}
\efig{Representative pulsar light curves in various wavelength regions
(from NASA GSFC).
\label{fig:curves}}

X rays and gamma rays are energetic, with thousands to billions of times more
energy per photon (light quantum) than visible light.  Even when a source is
very luminous at high energies (i.e., emitting a large amount of energy per
unit time), the number flux (number per unit time and area) of X~rays and
gamma~rays at Earth may be low.  Astronomers use instruments that can detect
and measure individual photons.  The resulting time series data are usually {\em
arrival time series}, sequences of precisely measured arrival times for
detected photons, $t_i$ ($i=1$ to $N$); photon energy and direction may also be
measured as ``marks'' on this point process.  For gamma-ray emission, the
flux is so small that the event rate is well below one event per period.  But
precise timing measurements spanning long time periods---hours to days---can
gather enough events to unambiguously identify pulsar signals, particularly
when multiple sets of observations spanning weeks or months (with large gaps)
are jointly analyzed.

In June 2008, NASA launched a new large space-based telescope tasked with
surveying the sky in gamma rays:  the {\em Fermi Gamma-Ray Space Telescope}. 
One of \Fermi's key scientific goals is to undertake a census of gamma-ray
pulsars (see Abdo et al.\ 2010 for the first \Fermi\ pulsar catalog).  This has
renewed interest in methods for analyzing arrival time
series data.  Here I will survey Bayesian work in this area dating from the
early 1990s that appears little-known outside of astronomy, and then describe
new directions for research motivated by \Fermi\ observations.

Since the photons originate from microscopic quantum mechanical processes at
different places in space, a Poisson point process (possibly non-homogeneous)
can very accurately model the data.  This is the foundation for both
frequentist and Bayesian approaches to periodic signal detection in these
data.  For bright X-ray sources, with many events detected per candidate
period, events may be binned in time, and standard periodogram techniques may
then be applied to the uniformly-spaced binned counts (the \pval\ calculation
is adjusted to account for the ``root--$n$'' standard deviation of the
counts).  We focus here instead on the low-flux case, where the data are too
sparse for binning to be useful, so they must be considered as a point
process.  This is the case for dim X-ray sources and all gamma-ray sources.

For most periodic signal detection problems with arrival time data ,
astronomers use frequentist methods inspired by the \pgram\ approach in the
additive noise setting described in \S~2:  one attempts to reject the null
model of constant rate by using a frequency-dependent test statistic,
calculating \pvals, and correcting for multiplicity.  A variety of statistics
have been advocated, but three dominate in practice (Lewis 1994 and Orford
2000 provide good overviews of the most-used methods).  All of them start by
folding the data modulo a trial period to produce a phase, $\phi_i$, for each
event in the interval $[0,2\pi]$; the statistics aim to measure departure from
uniformity over phase (i.e., they are statistics for detecting nonuniformity
of directional data on the circle).

First is the {\em Rayleigh statistic}, $R(\afreq)$, defined by
\beqnn
R^2 = \frac{1}{N}\left[\left(\sum_{i=1}^N \sin \phi_i \right)^2 +
      \left(\sum_{i=1}^N \cos \phi_i \right)^2\right].
\label{R-def}
\eeqnn
The quantity $2 R^2(\afreq)$ is called the
{\em Rayleigh power}.  It is the point process analog to the Schuster
\pgram\ of \ceqn{spgram-def}, and under the null, asymptotically
$2 R^2 \sim \chi^2_2$ (so $R^2$ follows an exponential distribution).
In practice, the Rayleigh statistic performs well for detecting signals
that have smooth light curves with a single peak per period.  As
Figure~1 reveals, this is not typically the case for high energy
emission from pulsars, so statistics are sought that have greater
power for more complicated shapes.

Taking a cue from the resemblence of $R(\afreq)$ to a Fourier magnitude,
the $\zsqm$ statistic sums power from $m$ harmonics (counting the
fundamental as $m=1$) of the Rayleigh power:
\beqnn
\zsqm = 2 \sum_{k=1}^m R^2(k\afreq).
\label{zsqm-def}
\eeqnn
Under the null, asymptotically $\zsqm \sim \chi^2_{2m}$.  The number of
harmonics, $m$, is usually set to a small integer value a priori ($m=2$ is
popular), though it is also possible to allow $m$ to adapt to the data.

The third commonly-used method is {\em $\chi^2$ epoch folding} (\efold).  For
every trial frequency, the folded phases are binned into $M$ equal-width phase
bins, and Pearson's $\chi^2$ is used to test consistency with the null
hypothesis of a constant phase distribution.  The number of bins is chosen a
priori.  The counts in each bin (for a chosen $\afreq$) will depend on the
origin of time; moving the origin will change the folded phases and shift
events between phase bins.  To account for this, the $\chi^2$ statistic may be
averaged over phase (Collura et al.\ 1987).  This alters its distribution
under the null; Collura et al.\ explore it via Monte Carlo simulation.


The $\zsqm$ and \efold\ statistics can be more sensitive to structured light
curves than the Rayleigh statistic, but with additional complexity in the form
of intractable distributions or the need to fix structure parameters (number
of harmonics or bins) a priori.

All of these statistics are simple to compute, and there are good reasons to
seek simplicity.  For a typical detectable X-ray pulsar, it may take
observations of duration $T \sim 10^4$ to $10^5$~s to gather a few thousand
photons; for a detectable gamma-ray pulsar, it may take a week or more of
integrated exposure time, so $T \sim 10^6$~s.  The Fourier spacing for such
data ranges from $\mu$Hz to $\sim 0.1$~mHz.  For a {\em targeted
search}---attempting to detect emission from a previously detected pulsar
(e.g., detected in radio waves)---the frequency uncertainty is typically
hundreds to thousands times greater than this Fourier spacing.  For a {\em
blind search}---attempting to discover a new pulsar---the number of
frequencies to search is orders of magnitude larger.  Pulsars are observed
with fundamental frequencies up to $\approx 700$~Hz (centrifugal forces would
destroy a neutron star rotating more rapidly than about a kilohertz).  The
non-sinusoidal shapes of pulsar light curves imply there may be significant
power in harmonics of the rotation frequency, at frequencies up to
$\fmax\approx 3000$~Hz.  The number of frequencies that must be examined is
then $\sim \fmax T$, which can be in the tens of millions for X-ray pulsars,
or the billions for gamma-ray pulsars.

In fact, the computational burden is significantly worse.  The energy emitted
by pulsars is drawn from the reservoir of rotational energy in the spinning
neutron star.  Thus, by conservation of energy, an isolated pulsar must be
spinning down (a pulsar in a binary system may instead spin {\em up}, if it is
close enough to its companion star to accrete mass carrying angular momentum).
 The pulsar frequency thus changes in time; a linear change, parameterized in
terms of the {\em frequency derivative} $\fdot$, describes most pulsars well,
though a few have higher derivatives that are measurable.  A pulsar search
must search over $\fdot$ values as well as frequency values.  The number of
$\fdot$ values to examine is determined by requiring that the frequency drift
across the data set, $\fdot T$, be smaller than the Fourier frequency spacing,
giving a number of $\fdot$ trials of $T^2\fdotmax$, with $\fdotmax \approx
10^{-10}$~Hz~s$^{-1}$ for known pulsars.  For a targeted search with the
shortest X-ray data sets, using a single $\fdot$ value (estimated from
previous observations) may suffice.  For blind searching for gamma-ray
pulsars, one may have to consider $\sim 10^3$ values of $\fdot$.  Clever use
of Fourier techniques, including tapered transforms, can reduce the burden
significantly (e.g., Atwood et al.\ 2006; Meinshausen et al.\ 2009).  Even so,
the number of effectively independent hypotheses in $(\freq,\fdot)$ space will
be thousands for targeted search, and many millions to a billion for blind
search.  This limits the complexity of detection statistics one may consider,
and requires that sampling distributions be estimated accurately far in their
tails.

We now consider Bayesian alternatives to the traditional tests, built using
time-domain models for a non-homogeneous Poisson point process with
time-dependent intensity (expected event rate per unit time)
$r(t)$.\footnote{The framework outlined here is presented in more detail
in an unpublished technical report (Loredo 1993); it was summarized in
Loredo (1992a), an abridged version of which appeared as Loredo (1992b).}
For periodic models, the parameters for $r(t)$ will include an amplitude, $A$;
the angular frequency, $\omega$; a phase (corresponding to defining an origin of
time), $\phi$; and one or more shape parameters, $\spars$, that parameterize the
light curve shape.  The likelihood function is,
\beqnn
\like(A,\afreq,\phi,\spars) = \exp\left[-\int_T dt\, r(t)\right]\;
      \prod_{i=1}^N r(t_i),
\label{like-def}
\eeqnn
written here with the parameter dependence implicit in
$r(t) = r(t; A,\afreq,\phi,\spars)$.

We will be comparing models for the signal, including a constant ``null''
model that will have only an amplitude parameter.  Since all models
share an amplitude parameter, it is helpful to define it in a way so
that a common prior may be assigned to $A$ across all models.  We write
the periodic model rate as,
\beqnn
r(t) = A \rho(\afreq t-\phi),
\label{r-def}
\eeqnn
where $\rho(\theta)$ is a periodic function with period $2\pi$,
and $A$ is defined to be the average rate,
\beqnn
A \equiv {1 \over P}\int_P dt\, r(t).
\label{A-def}
\eeqnn
(For a constant model, $r(t) = A$.)
This implies a normalization constraint on $\rho(\theta)$:
\beqnn
\int_0^{2\pi} d\theta\, \rho(\theta) = 2\pi,
\label{rho-norm}
\eeqnn
or, equivalently,
\beqnn
\int_{\per} dt\, \rho(\afreq t + \phi) = 1.
\label{rho-norm-2}
\eeqnn
That is, $\rho(\theta)$ is normalized as if $\rho(\theta)/2\pi$ were a
probability density in phase, or $\rho(\afreq t + \phi)$ were a probability
density in time (over one period).  With these definitions, the likelihood
function may be written,
\beqnn
\like(A,\afreq,\phi,\spars) = \left[A^N e^{-AT}\right] 
      \prod_i \rho(\afreq t_i - \phi).
\label{like-norm}
\eeqnn
Here we have presumed that $T$ spans many periods, so that the integral of the
rate over time in the exponent is well-approximated by $AT$.

Given an independent prior $\pi(A)$ for the amplitude, the marginal likelihood
for the frequency, phase, and shape is simply
\beqnn
\like(\afreq,\phi,\spars) \propto 
      \prod_i \rho(\afreq t_i - \phi),
\label{like-rho}
\eeqnn
where the constant of proportionality is the same for all models if a common
amplitude prior is used; it thus drops out of Bayes factors.\footnote{This
shared, independent prior assumption is a reasonable starting point for
analyzing individual systems, but deserves further consideration when
population modeling is a goal, since different physics may underly emission
from pulsars and non-pulsating neutron stars, and the expected amplitude of
pulsar emission likely depends on frequency (and other parameters). Since
amplitude and frequency are precisely estimated when a signal is detectable,
population modeling may be simplified in an empirical Bayes spirit by
inserting conditional prior factors, conditioned on the estimated amplitude
and frequency.}

To go forward, we now must specify models for $\rho(\theta)$, bearing
in mind the computational burden of $(\freq, \fdot)$ searching.  In
particular, since we will need to integrate the likelihood function
over parameter space (for evaluating marginals for estimation, and
marginal likelihood for model comparison), we seek models that allow
us to do as much integration {\em analytically} as possible.  Here
we focus on two complementary choices, one allowing analytical
phase marginalization, the other, a semiparametric model allowing
analytical shape parameter marginalization.

Since products of $\rho(\theta)$ appear in the likelihood, consider a
log-sinusoidal model, so that multiplication of rates leads to sums
of sinusoids in the likelihood.  Since $\rho(\theta)$ must be normalized,
this corresponds to taking $\rho$ proportional to a {\em von Mises
distribution},
\beqnn
\rho(\theta) = {1 \over I_0(\kappa)}\,e^{\kappa\cos(\theta)},
\label{rho-ls}
\eeqnn
where $I_0(\kappa)$ denotes the modified Bessel function of order 0.
This model has a single shape parameter, the {\em concentration parameter},
$\kappa$, that simultaneously controls the width of the peak in the
light curve, and the peak-to-trough ratio (or pulse fraction).

If we assign a uniform prior distribution for the phase (implied by time
translation invariance), a straightforward calculation gives the marginal
likelihood function for frequency and concentration:
\beqnn
\like(\afreq,\kappa) =  
       {I_0[\kappa R(\afreq)] \over [I_0(\kappa)]^N}.
\label{like-fk}
\eeqnn
The Rayleigh statistic arises as a kind of sufficient statistic for
estimation of frequency and concentration for a log-sinusoid model.
Interestingly, the $\kappa$ dependence depends only on the value
of $R$ and the sample size.  Using asymptotic properties of the
Bessel function one can show that, when there is potential evidence
for a signal at a particular frequency (amounting to $R > \sqrt{N}$),
the likelihood is approximately a gamma distribution in $\kappa$.
Also, the likelihood function strongly correlates $\afreq$ and $\kappa$,
so that the likelihood is largest at frequencies for which the
concentration would be estimated as large (which is intuitively
sensible).  A gamma distribution prior for $\kappa$ would be
asymptotically conjugate.

This is an interesting development because it opens the door to Bayesian
inference using computational tools already at hand for use of the Rayleigh
statistic (see Connors 1997 for a tutorial example calculation).  Bayesian
inferences for frequency, and for signal detection (via model comparison),
require integration of \ceqn{like-fk}\ over $\kappa$, but this is not a
significant complication.  A table of values of the integral may be
pre-computed at the start of the period search, as a function of $R$, and
interpolated for the final calculations.  Benefits of this Bayesian
counterpart to the Rayleigh test include simpler interpretation of results
(e.g., probability for a signal vs.\ a \pval), the possibility of integrating
the results into a multilevel model for population inferences, and the absence
of complex, sample-dependent corrections for non-independent test multiplicity
due to oversampling.

The complexity of the light curves in Figure~1 indicates that a model allowing
more structure than a single, smooth peak per period will be better able to
detect pulsars than the simple log-sinusoid model.  Ideally, one might consider
a richly flexible nonparametric model for $\rho(\theta)$, the overall model now
being semiparametric (with scalar parameters $\freq$, $\fdot$, and $\phi$). 
But the scale of the $(\freq,\fdot)$ search precludes use of a computationally
complex model. Inspired by the \efold\ method, Gregory and Loredo (1992; GL92)
consider a {\em piecewise constant shape} (PCS) model for $\rho(\theta)$, with
$\rho$ constant across $M$ equal-width phase bins.  Allowing $M$ to be
determined by the data makes this model semiparametric in spirit (in the
fashion of a sieve), if not formally nonparametric.


The PCS shape function may be written
\beqnn
\rho(\theta) = A M f_{j(\theta)},
  \quad \hbox{with}\quad
  j(\theta) = \left\lfloor 1+ M (\theta \bmod 2\pi) / 2\pi\right\rfloor,
\label{rho-steps}
\eeqnn
where the step parameters $\fvec = \{f_j\}$ specify the relative amplitudes of $M$ steps,
each of width $1/M$ period; with this parameterization, the step parameters
are constrained to be positive and to lie on the unit simplex, 
$\sum_j f_j = 1$.  The (marginal) likelihood function for angular frequency,
phase, and shape then has the form of a multinomial distribution:
\beqnn
\like(\afreq,\phi,\fvec) =  M^N \prod_{j=1}^M f_j^{n_j},
\label{like-steps}
\eeqnn
where $n_j=n_j(\omega,\phi)$ is the number of events whose times place them in
segment $j$ of the lightcurve, given the phase and frequency.  These numbers
correspond to the counts in bin $j$ in the EF method.

The appeal of the PCS model is the simple dependence on $\fvec$, which allows
analytic marginalization over shape if a conjugate prior is used.  GL92 adopted
a {\em flat shape prior}, $\pi(\fvec) = 1/M!$.  With this choice, the marginal
likelihood for frequency, phase, and $M$ is
\beqnn
\like(\omega,\phi)
  = {M^{N}\; (M-1)!\over (N+M-1)!}
    \left[n_1!\; n_2!\;\ldots n_M! \over N! \right].
\label{fs-mult}
\eeqnn
Only the term in brackets depends on $\omega$ and $\phi$.  It is just the
reciprocal of the multiplicity of the set of $n_j$ values---the number of ways
$N$ events can be distributed in $M$ bins with $n_j$ events in each bin.
Physicists know its logarithm as the configurational entropy of the $\{n_j\}$.
In fact, I devised this model specifically to obtain this result, formalizing
a clever intuition of Gregory's that entropy provides a measure of distance of
a binned distribution from a uniform distribution that could be superior to
the $\chi^2$ statistic used in \efold.  In a Bayesian setting, the reciprocal
multiplicity provides more than a simple test statistic; it enables
calculation of posterior probabilities for frequency, phase, and the number of
bins.  Further, by model averaging (over the choice of $M$, phase and
frequency), one can estimate the light curve shape without committing to a
particular binned representation.  A collection of pointwise estimates of
$\rho(\theta)$ vs. $\theta$ is {\em smooth} (albeit somewhat ``boxy''), though
considered as a function the estimate is outside the support of the model.

A drawback of the PCS model is that the phase parameter may not be
marginalized analytically.  Numerical quadrature must be used, which
makes the approach significantly more computationally burdensome
than the log-Fourier model (though not more burdensome than
phase-averaged \efold).

Figure~2 shows an example of the PCS model in action, from Gregory and Loredo
(1996; GL96).  These results use data from {\em ROSAT} satellite observations
of X-ray pulsar PSR~0540$-$693, located in the Large Magellanic Cloud, a small
irregular galaxy companion to the Milky Way.  This pulsar was first detected
in earlier data from the {\em Einstein Observatory} (Seward et al.\ 1984); it
is fast, with a period of $\approx 50$~ms. Later, less sensitive {\em ROSAT}
observations were undertaken to confirm the detection and improve the
estimated parameters, but the pulsar was not detectable using the Rayleigh
statistic (implemented via FFT).  Figure~2a shows the marginal likelihood for
the pulsar frequency for a 5-bin model, scaled to give the conditional odds in
favor of a periodic model over a constant model, were the frequency specified
a priori (and the constant model considered equally probable to the set of
models with $M=2$ to 10 a priori).  In fact, the prior measurements predicted
the frequency to lie within a range spanning $6\times 10^{-4}$~Hz (containing
about 144 Fourier frequencies for this data spanning $T=116,341$~s). 
Marginalizing over this range gives odds vs.\ $M$ as shown in Figure~2b. 
There is overwhelming evidence for the pulsar. Further results, including
light curve estimates and comparison with \efold, are in GL96.

\bfig{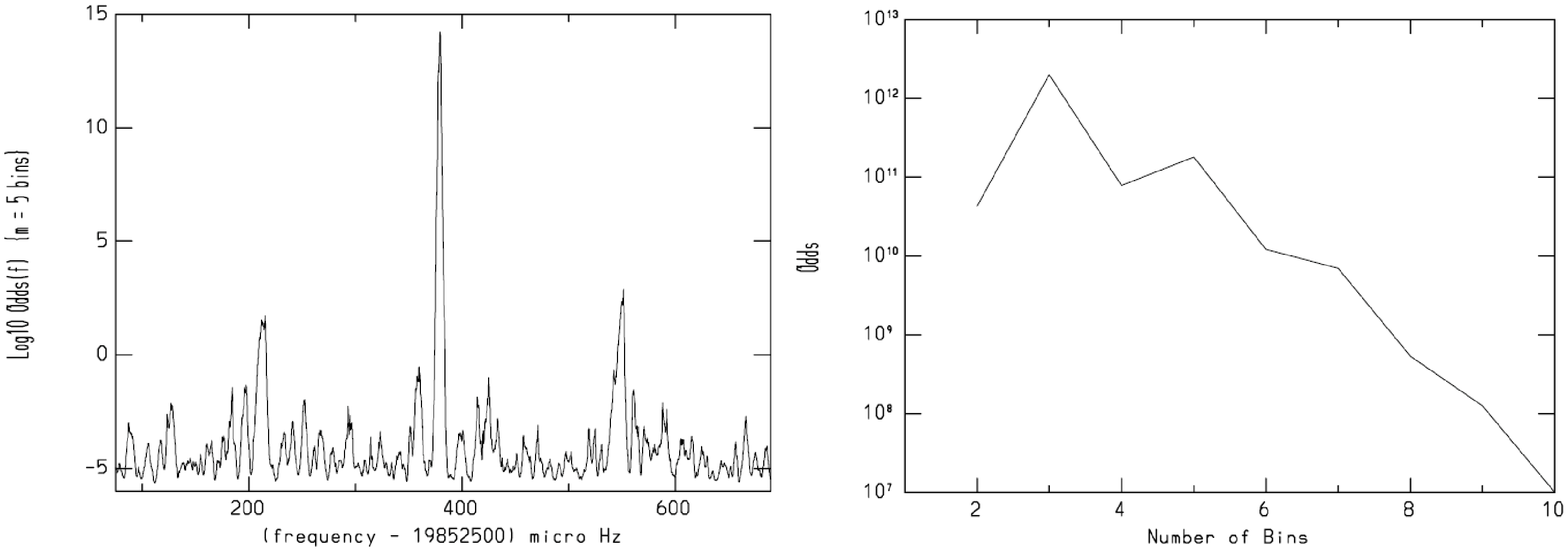}{120}
\efig{{\em (a)} Marginal likelihood for PSR~0540$-$693 frequency using
{\em ROSAT} data, for a 5-bin PCS model; likelihood scaled to indicate
the conditional (on frequency) odds favoring a periodic signal.
{\em (b)} Odds for a periodic model vs.\ a constant model, vs.\ 
number of bins.
\label{fig:LMC}}

A connection of the PCS model to \efold\ is worth highlighting.  Using
Stirling's approximation for the factorials in \ceqn{fs-mult}, one
can show that, for large numbers of counts in the bins,
\beqnn
\log \like(\omega,\phi)
  \approx {1\over 2}\chi^2 + {1\over 2}\sum_j \log n_j
          + C(M),
\label{chisqr}
\eeqnn
where $C(M)$ is a constant depending on $M$, and $\chi^2$ is the same
statistic used in the \efold\ method.  In fact, $\exp[-\chi^2/2]$ can be a
good approximation to the marginal likelihood for $\afreq$ and $\phi$. 
Despite this, in simulations the PCS model proves better able to detect weak
periodic signals than the phase-averaged $\chi^2$ statistic.  The reason
probably has less to do with failure of the approximation than with the fact
that, from a Bayesian viewpoint, the proper quantity to average over phase is
not $\chi^2$, but $\exp[-\chi^2/2]$.  Ad hoc averaging of $\chi^2$ to
eliminate the phase nuisance parameter essentially ``oversmooths'' in
comparison to a proper marginalization.

The launch of \Fermi\ has renewed interest in improving our capability to
detect weak periodic signals in arrival time series.  On the computational
front, important recent advances include the use of tapered transforms (Atwood
et al.\ 2006) and dynamic programming (Meinshausen et al.\ 2009) to accelerate
$(\freq,\fdot)$ exploration (in the context of Rayleigh and $\zsqm$
statistics).  Statistically, the most important recent development is the
introduction of likelihood-based score tests by Bickel et al.\ (2006, 2008). 
Inspired by the recent theoretical work on the limited power of omnibus tests
describe in \S~2, these tests seek high power in a family of models built with
a Fourier basis.  An interesting innovation of this approach is the use of
{\em averaging over frequency}, rather than maximizing, to account for
frequency uncertainty.  As in the \efold\ case, the averaging is of a quantity
that is roughly the logarithm of the marginal likelihood that would appear in
a Bayesian log-Fourier model.  It seems likely that a fully Bayesian treatment
of an analogous model could do better, though generalizing the log-sinusoid
model described above to include multiple harmonics is not trivial (Loredo
1993).

On the Bayesian front, a simple modification may improve the capability of the
PCS model.  Figures~3a,b show draws of shapes using the flat prior for $M=5$
and $M=30$; the shapes grow increasingly flat with growing $M$.  A better
prior would aim to stay variable as $M$ increases.  Consider the family of
conjugate symmetric Dirichlet priors (to keep the calculation analytic),
\beqnn
\pi(\fvec) \propto \delta\left(1-\sum_j f_j\right) \prod_j f_j^{\alpha-1}.
\label{dir-prior}
\eeqnn
One way to maintain variability is to make $\alpha$ depend on $M$ in a manner
that keeps the relative standard deviation of any particular $f_j$ constant with
$M$.  More fundamentally, we might seek to make the family of priors
{\em divisible}.  Both requirements point to the same fix:  take $\alpha =
C/M$, for some constant $C$.  Figure~3c shows samples from an $M=30$ prior
with $\alpha=2/M$ (the $M=2$ prior would be flat for this choice); variability
is restored. Informally, we might set $C$ a priori based on examination of
known light curves.  Alternately, inferring $C$ from the data, either
case-by-case or for populations (e.g., separately for X-ray and gamma-ray
pulsars), may provide useful insights into pulsar properties.  These avenues
are currently being explored.

\bfig{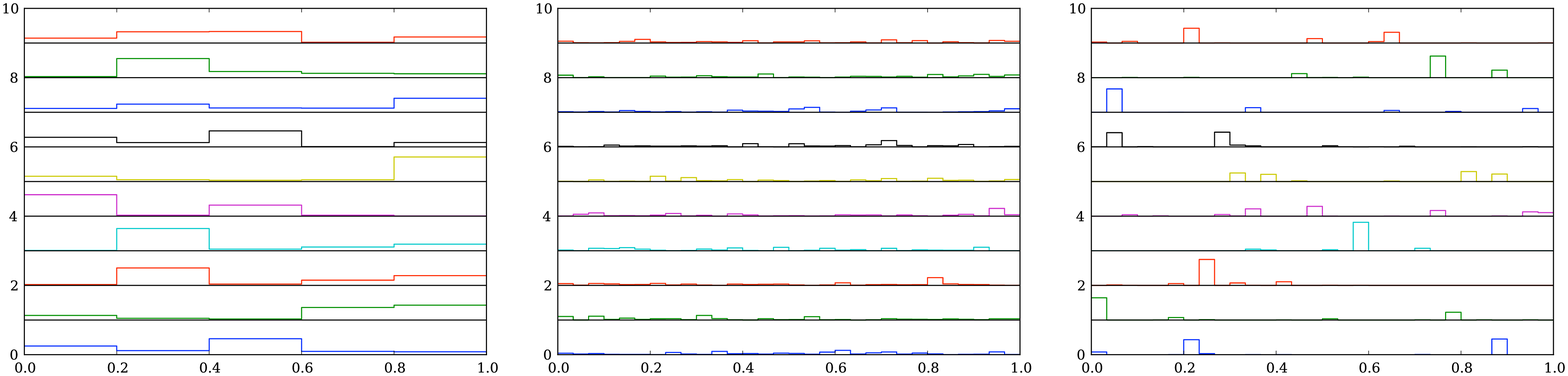}{120}
\efig{{\em (a, b, left)} 10 random samples (stacked) from a flat shape
distribution, for $M=5$ and $M=30$ bins.
{\em (c, right)} 10 random samples from a Dirichlet shape distribution
for $M=30$ bins, with $\alpha = 2/M$.
\label{fig:Dir}}

A possible approach for using more complex nonparametric Bayesian models may be
to use a computationally inexpensive method, like the log-sinusoid model or
the dynamic programming search algorithm of Bickel et al., for
a ``first pass'' analysis that identifies promising regions of $(\freq,\fdot)$
space.  The more complicated analysis would only be undertaken in the
resulting target regions.  However, the regions may still be large enough
to significantly constrain the complexity of nonparametric modeling.

We close this section with an observation about the apparently boring null
hypothesis, traditionally framed as a constant rate model, $r(t) = A$. 
It may be more accurate to frame it as a constant {\em shape}
model, $\rho(\theta) = 1$.  These do not quite amount to the same thing,
because in the shape description, we implicitly have a candidate period in
play, and we are asserting flatness of a ``per period'' or folded rate.  In
fact, few X-ray or gamma-ray sources have constant observed fluxes over
the duration of pulsar search observations.  Sources
often vary in luminosity in complex ways over time scales of hours and days. 
In some cases, the flux may vary because a survey instrument is not always
pointing directly at the source.  Although the rate as a function examined
over the full duration, $T$, may strongly vary, when folded over candidate
periods (always much smaller than $T$) and viewed vs.\ phase, it may be very
close to constant.  This is essentially an example of Poincar\'e's ``method of
arbitrary functions'' (e.g., Diaconis and Engel 1986).  Similar considerations
apply to periodic models:  models allowing period-to-period variability but
with a periodic expected rate can lead to the same likelihood function as
the strictly periodic models considered above.  These considerations remind us
that our hypotheses are always in some sense a caricature of reality, but that
in some cases we may be able to formally justify the caricature.

\section{Bayesian Inference and Design for Exoplanet\goodbreak Orbit Observations}

\begin{quote}\setlength{\leftskip}{3em}
Something there is more immortal even than the stars\ldots\\
Something that shall endure longer even than lustrous Jupiter\\
Longer than sun or any revolving satellite,\\
Or the radiant sisters the Pleiades.\\
\nobreak
\hbox{\hskip5em} ---{\em Walt Whitman}
\end{quote}

Ancient sky-watchers noted the complex movement of the planets with respect to
the fixed stars; in fact, ``planet'' derives from the Greek word for
``wanderer.''  Even before the heliocentric models of Copernicus, Galileo and
Kepler, this motion was attributed to {\em revolution} of the planets around
a host object, originally Earth, later the Sun.  By Newton's time, a more
sophisticated view emerged:  For an inertial observer (one experiencing no
measurable acceleration), the planets {\em and the Sun} appear to orbit around
their common center of mass.  The Sun is so much more massive than even the
most massive planet, Jupiter, that the center of mass of the solar
system---the {\em barycenter}---lies within the Sun (its offset from the Sun's
center is of order the solar radius, in a direction determined mostly by the
positions of Jupiter and Saturn).  The heliocentric descriptions of
Copernicus, Galileo and Kepler were approximations. Had they been able to make
precise observations of the solar system from a vantage point above the
ecliptic plane, they would have not only seen the planets whirling about in
large, elliptical, periodic orbits; they would have also seen the Sun
executing a complex, wobbling dance, albeit on a much smaller scale.

What the ancients could not see, and what modern instruments reveal, is that
some of the ``fixed'' visible stars are in fact wobbling on the sky, sketching
out small ellipses or more complex patterns similar to the Sun's unnoticed
wobble.  The largest motions arise from pairs of stars orbiting each other. 
But in the last 15 years, as a consequence of dramatic advances in
astronomers' ability to measure stellar motions, over 400 stars have been seen
to wobble in a manner indicating the presence of exoplanets.

To date, the most prolific technique for detecting exoplanets is the Doppler
radial velocity ``RV'' technique.  Rather than measuring the position of a
star on the sky versus time (which would require extraordinary angular
precision only now being achieved), this technique measures the {\em
line-of-sight velocity} of a star as a function of time---the toward-and-away
wobble rather than the side-to-side wobble.  This is possible using high
precision spectroscopic observations of lines in a star's spectrum; the
wavelengths of the lines shift very slightly in time due to the Doppler
effect. Radial velocities as small as a meter per second may be accurately
measured this way.


The resulting data comprise a time series of velocities measured with
additive noise, and irregularly spaced in time.  Figure~4 depicts a
typical data set and the currently dominant analysis method.
Figure~4a shows the velocity data; due to noise and the irregular
spacing, the kind of periodic time dependence expected from orbital
reflex motion is not visually evident, though it is clear something
is going on that noise alone cannot account for.  Figure~4b shows
a Lomb-Scargle periodogram of the data.  The resulting power spectrum
is very complex but has a clearly dominant peak.  The period corresponding
to the peak is used to initialize a $\chi^2$ minimization algorithm
that attempts to fit the data with a Keplerian orbit model, a strongly
nonlinear model describing the motion as periodic, planar, and elliptical.
Figure~4c shows the data folded with respect to the estimated period,
with the estimated Keplerian velocity curve; an impressive fit results.
For some systems, the residuals are large, and further periodic
components may be found by interative fitting of residuals, corresponding
to multiple-planet systems.  

\bfig{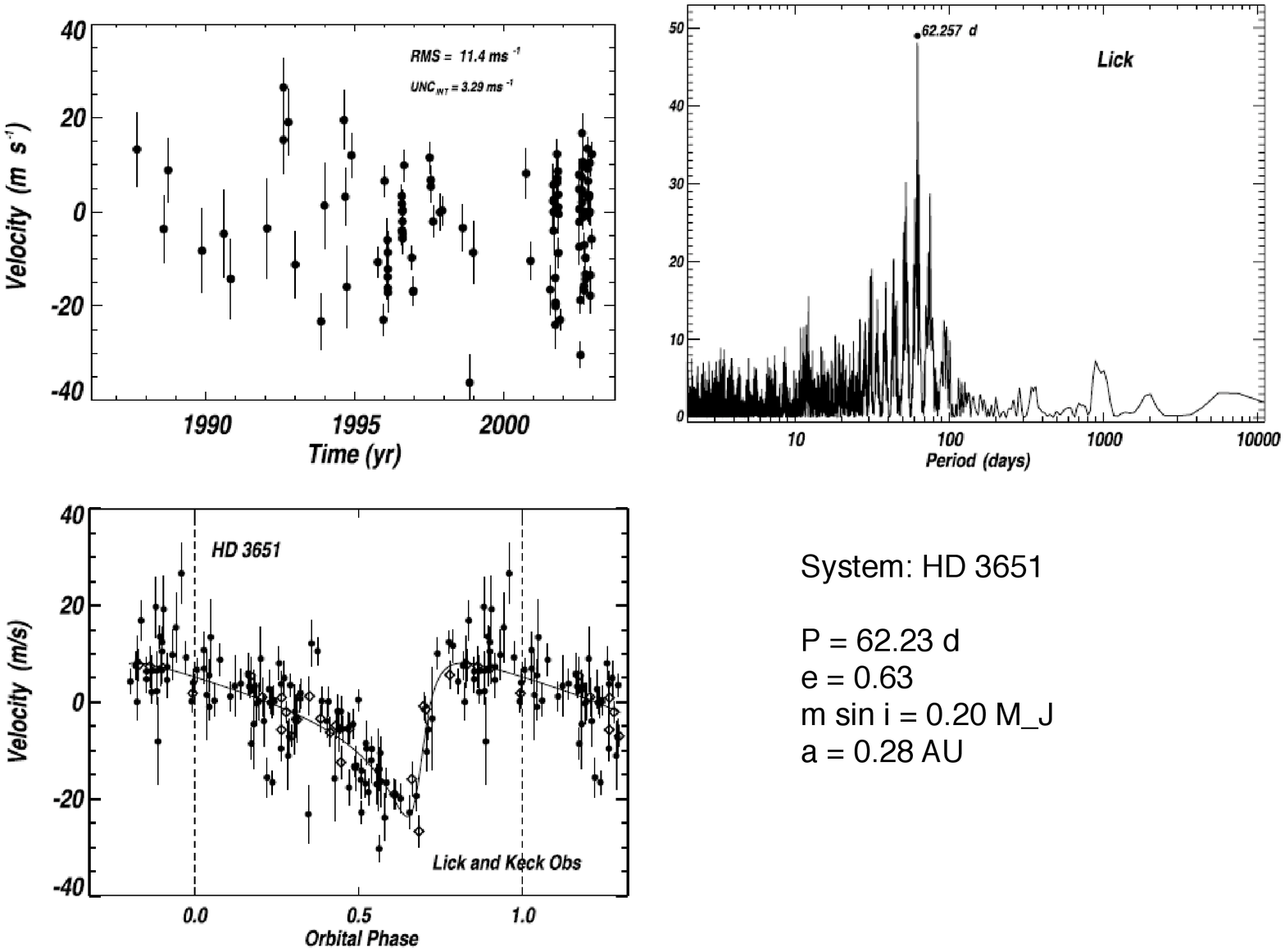}{120}
\efig{Depiction of the conventional RV data fitting process, based
on data from star HD~3651, from Fischer et al.\ (2003).
\label{fig:RVFit}}

This setting offers an interesting complement to pulsar data analysis. In both
problems, astronomers are searching for periodic signals.  But for planets,
there is a highly accurate parametric model for the signal.  Also, there is no
period derivative to contend with, and the number of frequencies to examine in
a blind search is typically thousands to hundreds of thousands, rather than
many millions or a billion (because the highest frequencies of interest are
far lower than in the pulsar case).  As a result, although periodograms are
part of the astronomer's tool kit in both settings, in other respects, the
data analysis methodologies differ greatly.

A number of challenges face astronomers analyzing exoplanet RV data with
conventional techniques.  The likelihood is highly multimodal, and in some
cases nonregular (e.g., for some orbital parameters, such as orbital
eccentricity, the likelihood is maximized on a boundary of parameter space).
The model is highly nonlinear.  As a consequence, Wilks's theorem is not
valid, and it becomes challenging to compute confidence regions from $\chi^2$
results.  Astronomers seek to use the orbital models to estimate derived
quantities such as planet masses, or to make predictions of future motion for
future observation; propagation of uncertainty in such calculations is
difficult.  As noted above, the LSP implicitly presumes a sinusoidal signal,
which corresponds to circular motion.  But many exoplanets are found to be in
eccentric orbits, so the LSP is suboptimal for exoplanet detection. These
challenges make it difficult to quantify uncertainty in marginal detections. 
As a result, only systems with unambiguous detections are announced, and the
implications of data from thousands of examined systems with no obvious
signals remains unquantified.  Finally, much of the interesting astrophysics
of exoplanet formation requires accurate inference of {\em population}
properties, but results produced by conventional methods make it challenging
to perform accurate population-level inferences.

Several investigators have independently turned to Bayesian methods to address
these challenges (Loredo and Chernoff 2000, 2003; Cumming 2004; Ford 2005;
Gregory 2005; Balan and Lahav 2008).  Here I will briefly describe ongoing
work I am pursuing in collaboration with my astronomer colleague David
Chernoff, and with statisticians Bin Liu, Merlise Clyde, and James Berger. The
most novel aspect of our work applies the theory of Bayesian experimental
design to the problem of {\em adaptive scheduling of observations} of
exoplanets.  Exoplanet observations use state-of-the-art instrumentation; the
observations are expensive, and observers compete for time on shared telescope
facilities.  It is important to optimize use of these resources.  This concern
will be even stronger for use of upcoming space-based facilities that will
enable measurement of the motion of the side-to-side positional wobble of
nearby stars.  Only relatively recently have simulation-based computational
techniques made it feasible to implement Bayesian experimental design with
nonlinear models (e.g., Clyde et al.\ 1995; M\"uller and Parmigiani 1995a,b;
M\"uller 1999).


Bayesian experimental design is an application of Bayesian decision theory,
and requires specification of a utility function to guide design.  Astronomers
have varied goals for exoplanet observations.  Some are interested in
detecting individual systems; others seek systems of a particular type (e.g.,
with Earth-like planets) and may want to accurately predict planet positions
for future observations (e.g., of transits of a planet across the disc of its
host star); others may be interested in population properties.  No single,
tractable utility function can directly target all of these needs.  We thus
adopt an information-based utility function, as described by Lindley (1956,
1972) and Bernardo (1979), as a kind of ``general purpose'' utility.


As a simple example, consider observation of an exoplanet system with a single
detected planet, with the goal of refining the posterior distribution for
the orbital parameters, $\theta$.  Denote the currently available data
by $D$, and let $M_1$ denote the information specifying the single-planet
Keplerian orbit model.  The current posterior distribution for the orbital
parameters is then $p(\theta|D,M_1)$ (we will suppress $M_1$ for the time
being).  For an experiment, $e$, producing
future data $d_e$, the updated posterior will be $p(\theta|d_e,D)$;
here $e$ labels the action space (e.g., the time for a future observation),
and $d_e$ is the associated (uncertain) outcome.  We take the utility
to be the information in the updated posterior, quantified by
the negative Shannon entropy,
\beqnn
\Info(e,d_e) 
 = \int d\theta\, p(\theta|d_e,D)\, \log\left[p(\theta|d_e,D)\right] 
\label{util-shannon}
\eeqnn
(using the Kullback-Leibler divergence between the original and updated
posterior produces the same results; we use the Shannon entropy here
for simplicity).  The optimal experiment maximizes the expected information,
calculated by averaging over the uncertain value of $d_e$;
\beqnn
\einfo(e) = \int dd_e\, p(d_e|D) \, \Info(e,d_e),
\label{einfo}
\eeqnn
where the predictive distribution for the future data is
\beqnn
p(d_e|D) = \int d\theta\, p(\theta|D) p(d_e|\theta).
\label{predict}
\eeqnn

Calculating the expected information in \ceqn{einfo}\ requires evaluating a
triply-nested set of integrals (two over the parameter space, and one over the
future sample space); we must then optimize this over $e$.  This is a
formidable calculation.  But a significant simplification is available in some
settings.  Sebastiani and Wynn (2000) point out that when the information in
the future sampling distribution, $p(d_e|\theta)$, is independent of the
choice of hypothesis (i.e., the parameters, $\theta$), the expected
information simplifies;
\beqnn
\einfo(e) = C - \int dd_e\, p(d_e|\theta) \log[p(d_e|\theta)],
\label{maxent}
\eeqnn
where $C$ is a constant (measuring the $e$-independent information in the
prior and the sampling distribution).  The integral (including the minus sign)
is the Shannon entropy in the predictive distribution.  Thus the experiment
that maximizes the expected information is the one for which the predictive
distribution has minimum information, or maximum entropy.  The strategy of
sampling in this optimal way is called {\em maximum entropy sampling}
(MaxEnt sampling). 
Colloquially, this strategy says we will learn the most by sampling where we
know the least, an appealingly intuitive criterion.

As a simplified example, consider an RV data model with measurements given
by
\beqnn
d_i = V(t_i; \tau, e, K) + \eps_i,
\label{d=Ve}
\eeqnn
where $\eps_i$ denotes zero-mean Gaussian noise terms with known variance
$\sigma^2$, and $V(t_i; \tau, e, K)$ gives the Keplerian velocity along the
line of site as a function of time $t_i$ and of the orbital parameters $\per$
(period), $e$ (eccentricity), and $K$ (velocity amplitude).  For simplicity
two additional parameters required in an accurate model are held fixed: a
parameter describing the orbit orientation, and a parameter specifying the
origin of time.  The velocity function is strongly nonlinear in all variables
except $K$ (its calculation requires solving a famous transcendental equation,
the Kepler equation; see Danby 1992 for details).  Our goal is to learn about
the parameters $\tau$, $e$ and $K$.

Figure~5 shows results from a typical simulation iterating an
observation-inference-design cycle a few times. Figure~5a shows simulated data
from a hypothetical ``setup'' observation stage.  Observations were made at 10
equispaced times; the curve shows the true orbit with typical exoplanet
parameters ($\tau=800$~d, $e=0.5$, $K=50$~ms$^{-1}$), and the noise
distribution is Gaussian with zero mean and $\sigma=8$~m~s$^{-1}$.  Figure~5b
shows some results from the inference stage using these data.  Shown are 100
samples from the marginal posterior density for $\tau$ and $e$ (obtained with
a simple but inefficient accept/reject algorithm). There is significant
uncertainty that would not be well approximated by a Gaussian (even
correlated).  Figure~5c illustrates the design stage.  The thin curves display
the uncertainty in the predictive distribution as a function of sample time;
they show the $V(t)$ curves associated with 15 of the parameter samples from
the inference stage.  The spread among these curves at a particular time
displays the uncertainty in the predictive distribution at that time.  A Monte
Carlo calculation of the expected information vs.\ $t$ (using all 100 samples)
is plotted as the thick curve (right axis, in bits, offset so the minimum is
at 0 bits).  The curve peaks at $t=1925$~d, the time used for observing in the
next cycle.

\bfig{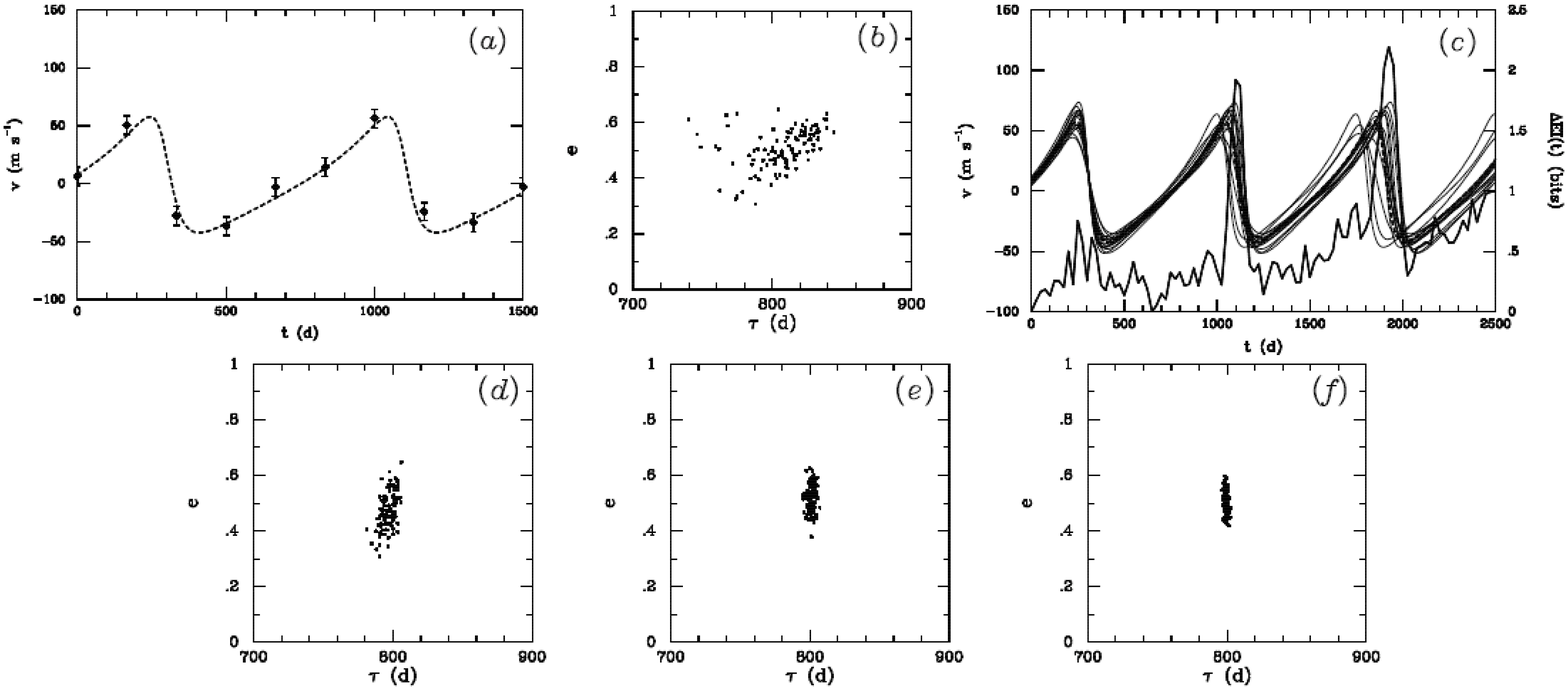}{120}
\efig{Initial observations {\em (a, top left)}, interim inferences
{\em (b, top middle)}, and design stage {\em (c, top right)} for a
simulated observation-inference-design cycle implementing adaptive
design for a simplified eccentric exoplanet model.  
{\em (d--f, bottom)} Evolution of inferences in subsequent cycles.
\label{fig:BAE}}

Figure~5d shows interim results from the inference stage of the next cycle
after making a single simulated observation at the optimal time. The period
uncertainty has decreased by more than a factor of two, and the product of the
posterior standard deviations of all three parameters (a crude measure of
``posterior volume'') has decreased by a factor $\approx 5.8$; this was
accomplished by incorporating the information {\em from a single well-chosen
datum}.  Figures~5e,f show similar results from the next two cycles.  The
posterior volume continues to decrease much more rapidly than one would expect
from the random-sampling ``$\sqrt{N}$ rule'' (by factors of $\approx 3.9$ and
$1.8$).

To implement this approach with the full Keplerian RV model requires a
nontrivial posterior sampling algorithm.  One pipeline we have developed is
inspired by the conventional LSP$+\chi^2$ technique.  As a starting point, we
use the fact that the Keplerian velocity model is a {\em separable} nonlinear
model, which may be reparameterized as a linear superposition of two nonlinear
components.  We can analytically marginalize over the two linear parameters,
producing a marginal likelihood for three nonlinear parameters:  $\per$, $e$,
and an origin-of-time parameter, $\mu_0$ (an angle denoting the orbit
orientation at $t=0$).  We eliminate $e$ and $\mu_0$, either by crude
quadrature, or by using heuristics from Fourier analysis of the Keplerian
model to estimate values from a simple harmonic fit to the data.  This
produces an approximate marginal likelihood for period that we call a {\em
Kepler periodogram} (K-gram).  It plays the role of the LSP in the
conventional analysis, but accounts for orbital eccentricity.

The K-gram (multiplied by a log-flat prior in period) is an approximate
marginal density for the period.  Rather than use a periodogram peak to
initialize a $\chi^2$ parameter fit, we draw $\sim 10$ to 20 samples from the
K-gram to define an initial {\em population} of candidate orbits.  Finally, we
evolve the population using a population-based adaptive MCMC algorithm.  Our
current pipeline uses the differential evolution MCMC algorithm of Ter Braak
(2006).  When applied to simulated and real data for systems with a single,
well-detected exoplanet, this pipeline produces posterior samples much more
efficiently than other recently-developed algorithms (e.g., the random walk
Metropolis algorithm of Ford 2005, or the parallel tempering algorithm of
Gregory 2005).  The success of the algorithm appears due to the ``smart start''
provided by the K-gram, and the adaptivity of population-based MCMC.

However, this pipeline has limitations that have led us to explore more
thoroughgoing departures from existing algorithms.  The first limitation
is that when there is significant multimodality (i.e., more than one
mode with significant posterior probability), our population-based
sampler explores parameter space much less efficiently due to the difficulty 
of swapping between modes.

The second limitation is more fundamental.  So far, we have focused on
adaptive design for parameter estimation, presuming the stellar target
is known to host a star.  In fact, initially we will not know whether a
star hosts a planet or not; we initially need to optimize for {\em detection}
(i.e., model comparison), not estimation.  Even after a planet is detected,
while we would like future observations to improve the orbital parameter
estimates, we would also like the observational design to consider the
possibility that an additional planet may be present.

To pursue more general design goals, we introduce a set of models,
$M_k$, with $k$ planets ($k=0$ to a few), with associated parameter
spaces $\theta_k$.  Write the joint posterior for the models and
their parameters as
\beqnn
p(M_k,\theta_k|d_e, D)
  &=& p(M_k|d_e, D) p(\theta_k|d_e, D,M_k) \;\equiv\; p_k\; q_k(\theta_k),
\label{joint}
\eeqnn
where $p_k$ is the posterior probability for $M_k$, and $q_k(\theta_k)$
is the posterior density for the parameters of model $M_k$.  Then the
information in the joint posterior is,
\beqnn
\Info[M_k,\theta_k|D]
  &=& \sum_k \int d\theta_k\, p_k q_k(\theta_k) \log [p_k q_k(\theta_k)]\\
  &=& \sum_k p_k \log p_k + \sum_k p_k \int d\theta_k\, q_k(\theta_k) \log
q_k(\theta_k).
\label{info-joint}
\eeqnn
The first sum in \ceqn{info-joint}\ is the information (negative entropy) in
the posterior over the models; the second sum averages the information in the
various posterior densities, weighted by the model probabilities.  Once the
data begin to focus on a particular model (so one of the $p_k$ values
approaches unity and the others approach zero), the first term will nearly
vanish, and the sum comprising the second term will be dominated by the term
quantifying the information in the posterior density for the best model.  That
is, the parameter estimation case described above is recovered.  When model
uncertainty is signficant, the first term plays a significant role, allowing
model uncertainty to drive the design.  This utility thus naturally moves
between optimizing for detection and for parameter estimation. We have found
that Borth (1975) derived essentially the same criterion, dubbed a {\em total
entropy criterion}, though it has gone unused for decades, presumably because
the required calculations are challenging.


Three features make use of this more general criterion significantly more
challenging than MaxEnt sampling.  First, model probabilities are needed,
requiring calculation of marginal likelihoods (MLs) for the models.  MCMC
methods do not directly estimate MLs; they must be supplemented with other
techniques, or MCMC must be abandoned for another approach.  Second, the
condition leading to the MaxEnt simplification in the parameter estimation
case---that the entropy in the predictive distribution not depend on the
choice of hypothesis---does not hold when the hypothesis space includes
composite hypotheses (marginalization over rival models' parameter
spaces breaks the condition).

Finally, for adaptive design for parameter estimation above, we adopted
a greedy algorithm, optimizing one step ahead.  For model choice,
it is typically the case that non-greedy designs significantly out-perform
greedy designs (more so than for parameter estimation).  This significantly
complicates the optimization step.

Motivated by these challenges, we have developed an alternative computational
approach that aims to calculate marginal likelihoods directly, producing
posterior samples as a byproduct:  {\em annealing
adaptive importance sampling} (AAIS).  This algorithm anneals a target
distribution (prior times likelihood for a particular model), and
adapts an importance sampler built out of a mixture of multivariate
Student-$t$ distributions to the sequence of annealed targets, using
techniques from sequential Monte Carlo.  The number of components in
the mixture adapts via birth, death, merge and split operations; the
parameters of each component adapt via expectation-maximization algorithm
steps.  The algorithm currently works well on several published data
sets with multimodal posteriors and either one or two planets.
A forthcoming publication (Liu et al.\ 2011) provides details.

\section{Perspective}  

I have highlighted here only two among many areas in astronomy where
astronomers study periodic phenomena. So far Bayesian methods are relatively
new for such problems. I know of only two other applications where astronomers
are studying periodic phenomena with Bayesian methods:  Berger et al.\ (2003)
address nonparametric modeling of Cepheid variable stars that are used to
measure distances to nearby galaxies (via correlation between luminosity and
period); and Brewer et al.\ (see White et al.\ 2010 and references therein)
address detection and estimation of low-amplitude, nearly-periodic
oscillations in stellar luminosities (asteroseismology).

Broadening the perspective beyond periodic phenomena, astronomy is on the
verge of a revolution in the amount of time-domain data available.  Within a
decade, what was once the science of the fixed stars will become a thoroughly
time-domain science.  While much time-domain astronomy to date has come from
targeted observations, upcoming large-scale surveys will soon produce
``whole-sky time-lapse movies'' with many-epoch multi-color observations of
hundreds of millions of sources.  The prime example is the {\em Large Synoptic
Survey Telescope}, which will begin producing such data in 2019.  Hopefully the
vastness and richness of the new data will encourage further development of
Bayesian tools for exploring the dynamic sky.


In this decade marking dramatic growth in the importance and public visibility
of time-domain astronomy, it is perhaps not surprising to find contemporary
writers relinquishing stars as symbols of steadfastness; they are instead
symbols of enduring mystery.  In her poem, ``Stars'' (Manfred 2008),
Wisconsin-based poet Freya Manfred depicts a moment of exasperation at life in
a mercurial world, with the poet finding herself ``past hanging on.'' 
One thing is able to distract her from the vagaries of daily life---not the
illusory steadfastness of the once fixed stars, but the enigma
of the pulsating sky:


\begin{quote}
But I don't care about your birthday, or Christmas, or lover's lane,\\
or even you, not as much as I pretend. Ah, I was about to say,\\
``I don't care about the stars'' --- but I had to stop my pen.

Sometimes, out in the silent black Wisconsin countryside\\
I glance up and see everything that's not on earth, glowing, pulsing,\\
each star so close to the next and yet so far away.

Oh, the stars. In lines and curves, with fainter, more mysterious\\
designs beyond, and again, beyond.  The longer I look, the more I see,\\
and the more I see, the deeper the universe grows.
\end{quote}

\bref

\rr Abdo, A. A. et al.\ (2010).
The First {\em Fermi} Large Area Telescope Catalog of Gamma-Ray Pulsars. 
{\em Astrophys. J. Supp. Ser.}  {\bf 187},  460--494.

\rr Atwood, W. B., Ziegler, M., Johnson, R. P. and Baughman, B. M. (2006).
A Time-differencing Technique for Detecting Radio-quiet Gamma-Ray Pulsars.
{\em Astrophys. J.}  {\bf 652},  L49--L52.

\rr Balan, S. T. and Lahav, O. (2008).
EXOFIT: orbital parameters of extrasolar planets from radial velocities. 
{\em Mon. Not. Roy. Ast. Soc.} {\bf 394}, 1936--1944.

\rr Berger, J. O., Jefferys, W. H., M\"uller, P. and Barnes, T. G. (2003).
Bayesian model selection and analysis for Cepheid star oscillations. 
{\em Statistical challenges in astronomy} (E. D. Feigelson and G. J. Babu, eds.)
New York: Springer, 71--88.

\rr Bernardo, J. M. (1979).
Expected information as expected utility.   \as{7}, 686--690.

\rr Bickel, P., Kleijn, B., Rice, J. (2008).
Weighted Tests for Detecting Periodicity in Photon Arrival Times.
{\em Astrophys. J.} {\bf 685}, 384--389.

\rr Bickel, P. J., Ritov, Y., Stoker, T. M. (2006).
Tailor-made tests for goodness of fit to semiparametric hypotheses.
\as{34}, 721--741.

\rr Borth, D. M. (1975).
A total entropy criterion for the dual problem of model determination
and parameter estimation.  \jrssb{37}, 77--87.

\rr Bretthorst, G. L. (1988).
{\em Bayesian Spectrum Analysis and Parameter Estimation}.  Berlin: Springer-Verlag.

\rr Bretthorst, G. L. (2001).
Nonuniform sampling: Bandwidth and aliasing. 
{\em Bayesian Inference and Maximum Entropy Methods in Science and Engineering}
(J. Rychert, G. Erickson and C. R. Smith, eds.).  New York:  American Inst.\ of
Physics, pp. 1--28.

\rr Clyde, M., M\"uller, P. and Parmigiani, G. (1995).
Exploring expected utility surfaces by markov chains. 
ISDS Discussion Paper 95-39, Duke University.

\rr Collura, A., Maggio, A., Sciortino, S., Serio, S., Vaiana, G. S. and
osner, R. (1987).
Variability analysis in low count rate sources. 
{\em Astrophys. J.} {\bf 315}, 340--348.

\rr Connors, A. (1997). 
Periodic analysis of time series data as an exemplar of Bayesian methods.
{\em Data Analysis in Astronomy}
(V. Di Gesu, M. J. B. Duff, A. Heck, M. C. Maccarone, L. Scarsi and H.
U. Zimmerman, eds.). Singapore: World Scientific Press, 251--260.


\rr Cumming, A. (2004).
Detectability of extrasolar planets in radial velocity surveys. 
{\em Mon. Not. Roy. Ast. Soc.} {\bf 354}, 1165--1176.

\rr Danby, J. M. A. (1992).
{\em Fundamentals of Celestial Mechanics}.
Richmond, VA: William-Bell, Inc.

\rr Diaconis, P. and Engel (1986).
Comment on `Application of Poisson's Work'. 
\stsc{1}, 171--174.

\rr Fischer, D. A., Butler, R. P., Marcy, G. W., Vogt, S. S., Henry, G. W.
(2003).
A sub-Saturn mass planet orbiting HD~3651.
{\em Astrophys. J.} {\bf 590}, 1081--1087.

\rr Freedman, D. (2009).
Diagnostics cannot have much power against general alternatives.
{\em Int. J. Forecasting} {\bf 25}, 833--839.

\rr Ford, E. (2005).
Quantifying the uncertainty in the orbits of extrasolar planets. 
{\em Astronomical J.} {\bf 129}, 1706--1717.

\rr Gregory, P. (2005).
A bayesian analysis of extrasolar planet data for HD 73526. 
{\em Astrophys. J.} {\bf 631}, 1198--1214.

\rr Gregory, P. and Loredo, T. J. (1992).
A new method for the detection of a periodic signal of unknown shape and
period. 
{\em Astrophys. J.} {\bf 398}, 148--168.

\rr Gregory, P. and Loredo, T. J. (1996).
Bayesian periodic signal detection: Analysis of ROSAT observations of
PSR~0540$-$693.
{\em Astrophys. J.} {\bf 473}, 1059--1056.

\rr Hopkins, A. M. et al. (2002).
A new source detection algorithm using the false-discovery rate. 
{\em Astron. J.} {\bf 123}, pp. 1086--1094.

\rr Janssen, A. (2000).
Global power functions of goodness of fit tests.
\as{28}, 239--253.

\rr Jaynes, E. T. (1987).
Bayesian spectrum and chirp analysis. 
{\em Maximum Entropy and Bayesian Spectral Analysis and Estimation
Problems} (C. R. Smith and G. J. Erickson, eds.) Dordrecht: D.~Reidel,
1.

\rr Lehmann, E. L., Romano, J. P. (2005).
{\em Testing Statistical Hypotheses}. \springerny.

\rr Lewis, D. A. (1994).
Weak periodic signals in point process data. 
{\em Statistical methods for physical science, Methods of Experimental Physics Vol.~28}
(J. L. Stanford and S. B. Vardeman, eds.) San Diego: Academic Press,
349--373.

\rr Lindley, D. V. (1956).
On a measure of the information provided by an experiment. 
\ams{27}, 986--1005.

\rr Lindley, D. V. (1972).
{\em Bayesian Statistics---A Review}. Montpelier: SIAM/Capital City Press.

\rr Liu, B., Clyde, M., Berger, J. O., Loredo, T. J., Chernoff, D. C. (2010).
An adaptive annealed importance sampling method for calculating marginal
likelihoods with application to bayesian exoplanet data analysis. 
In preparation.

\rr Lomb, N. R. (1976).
Least-squares frequency analysis of unequally spaced data.
{\em Astrophys. Sp. Sci.} {\bf 39}, 447--462.

\rr Loredo, T. J. (1992a).
The promise of Bayesian inference for astrophysics.
(Unabridged version of Loredo 1992b)
{\tt http://citeseerx.ist.psu.edu/viewdoc/summary?doi=10.1.1.56.1842}, 1--49.

\rr Loredo, T. J. (1992b).
The promise of Bayesian inference for astrophysics. 
{\em Statistical Challenges in Modern Astronomy}
(E.~Feigelson and G.~J.~Babu, eds.) New York: Springer-Verlag, 275--306 
(with discussion).

\rr Loredo, T. J. (1993).
Bayesian inference with log-Fourier arrival time
models and event location data.  Technical report.  
{\tt http://www.astro.cornell.edu/staff/loredo/}
 
\rr Loredo, T. J. and Chernoff, D. C. (2000).
Bayesian methodology for the space interferometry mission. 
{\em Bull. Am. Astron. Soc.} {\bf 32}, 767.

\rr Loredo, T. J. and Chernoff, D. C. (2003).
Bayesian adaptive exploration.  
{\em Statistical challenges in astronomy}
(E. D. Feigelson and G. J. Babu, eds.) New York: Springer,
57--70.

\rr Loredo, T. J. (2004).
Bayesian adaptive exploration. 
{\em 23rd International Workshop on Bayesian Inference and Maximum Entropy
Methods in Science and Engineering} (G. J. Erickson and
Y. Zhai, eds.)
New York:  AIP Conference Proceedings {bf 707}, 330--346.

\rr Manfred, F. (2008).
{\em Swimming with a Hundred Year Old Snapping Turtle}. 
Northfield, MN: Red Dragonfly Press.

\rr Meinshausen, N., Bickel, P. and Rice, J. (2009).
Efficient blind search: Optimal power of detection under computational cost
constraints. 
{\em Ann. Appl. Stat.} {\bf 3}, 38--60.

\rr M\"uller, P. and Parmigiani, G. (1995a).
Numerical evaluation of information theoretic measures. 
{\em Bayesian Statistics and Econometrics: Essays in Honor of A. Zellner}
(D. A. Berry, K. M.
Chaloner and J. F. Geweke, eds.) New York: Wiley, 397--406.

\rr M\"uller, P. and Parmigiani, G. (1995b).
Optimal design via curve fitting of monte carlo experiments.
\jasa{90}, 1322--1330.

\rr M\"uller, P. (1999).
Simulation based optimal design. 
\val6, 459--474.

\rr Orford, K. J. (2000).
The analysis of cosmic ray data. 
{\em J. Phys. G: Nucl. Part. Phys.} {\bf 26}, R1--R26.

\rr Scargle, J. D. (1982).
Studies in astronomical time series analysis. II - Statistical aspects of
spectral analysis of unevenly spaced data. 
{\em Astrophys. J.} {\bf 263}, 835--853.

\rr Schuster, A. (1898).
On the investigation of hidden periodicities with application to a supposed 26
day period of meteorological phenomena.
{\em Terrestrial Magnetism and Atmospheric Electricity} {\bf 3}, 13--41

\rr Sebastiani, P. and Wynn, H. P. (2000).
Maximum Entropy Sampling and Optimal Bayesian Experimental Design. 
\jrssb{62}, 145--157.

\rr Ter Braak, C. J. F. (2006).
A Markov chain Monte Carlo version of the genetic algorithm
differential Evolution: Easy Bayesian computing for real
parameter spaces.  \sc{16}, 239--249.

\rr 	
White, T. R., Brewer, B. J., Bedding, T. R., Stello, D. and Kjeldsen, H. (2010).
A comparison of Bayesian and Fourier methods for frequency determination in
asteroseismology.
{\em Comm. in Asteroseismology} {\bf 161}, 39--53.

\eref

\newpage

\centerline{\large Response to discussion by Peter M\"uller}
\bigskip

As a non-statistician interloper of sorts, I am grateful to the organizers for
the privilege of being invited to participate in this last Valencia meeting,
and for assigning me so effective (and gracious) a discussant.  Prof.\ Peter
M\"uller
presents a number of useful new ideas and clarifying questions in his
deceptively short discussion.  I will touch on a selection of his points in
this response; limits of space provide me a convenient excuse for postponing
the address of other important points for another forum where I may have
``pages enough and time'' to explore M\"uller's suggestions more fully.

For the pulsar detection problem, M\"uller suggests changing the prior to
a symmetric Dirichlet prior with a small exponent in order to favor light
curve shapes with spikes.  With a similar motivation, in the paper I proposed
adopting a divisible Dirichlet prior, say with $\alpha = 2/M$ for the $M$-bin
shape model.  This becomes a small-$\alpha$ prior once $M$ is larger than
a few.  Preliminary calculations indicate this is a promising direction,
but not entirely satisfactory.  Figure~\ref{fig:Divis}\ shows, as a function
of number of bins, the Bayes factor for a model using the divisible prior
versus one using the Gregory \& Loredo flat prior, for three representative
types of data.  For data distributing events uniformly across the bins, the
(red)
squares show that adopting the divisible prior allows one to more securely
reject periodic models.  For data placing all events in a single-bin pulse,
the (blue) diamonds show that the divisible prior results in dramatically
increased sensitivity to pulsations.  However, as is evident in Figure~1 in
the paper, gamma-ray pulsations typically ride on top of a constant background
component.  Adding such a component to the single-bin pulse data (at about 9\%
of the pulse level) produces the Bayes factors indicated by the (green) circles;
these indicate {\em less} sensitivity to pulsations with the divisible prior
than with the flat prior.  Small-$\alpha$ priors put prior mass on truly
spiked signals, with all events in very few bins.  This preference has to
be tempered in order to realistically model pulsar light curves with a
background component.  I am exploring how to achieve this, following some
of M\"uller's leads.

\bfig{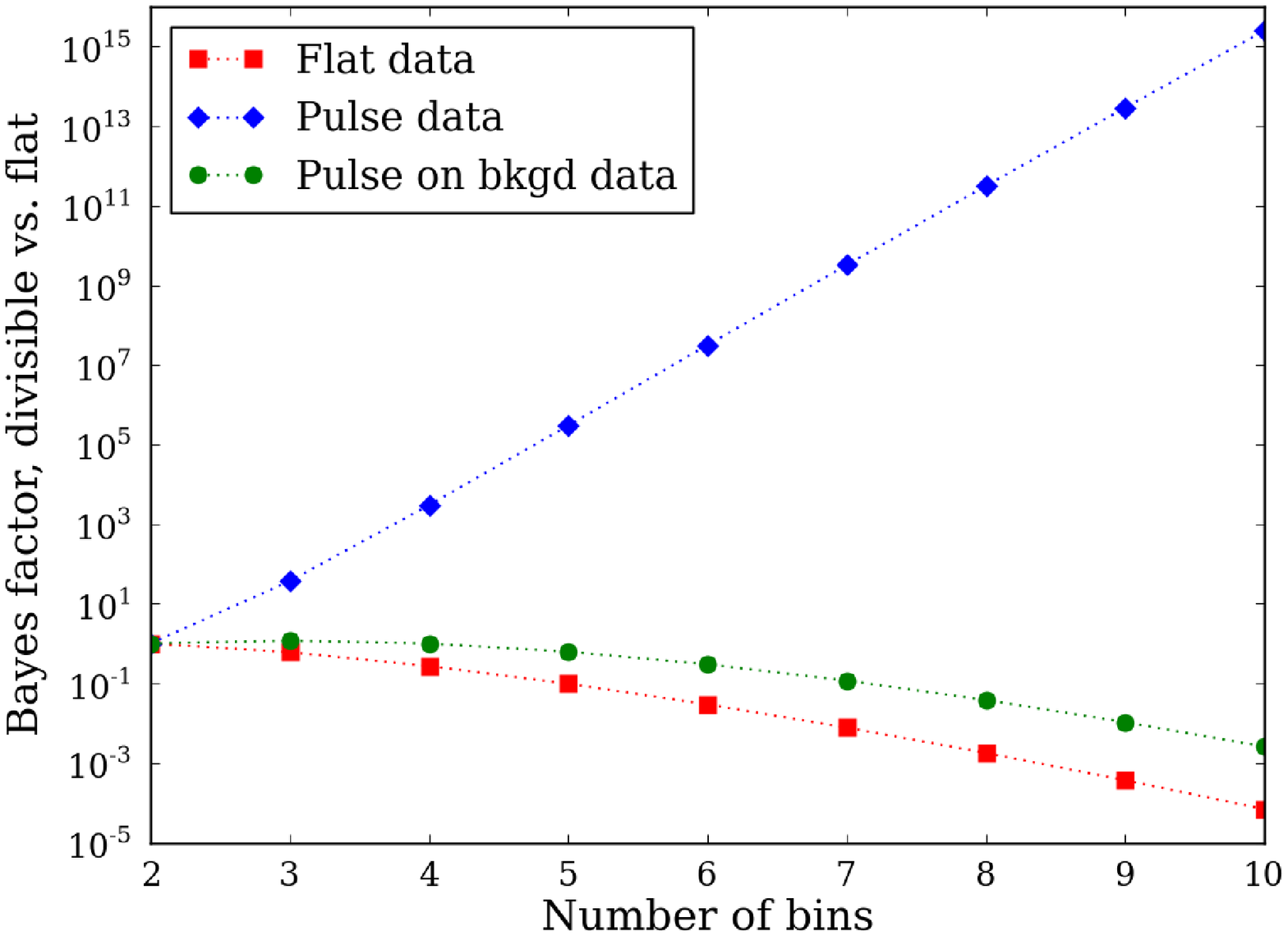}{90}
\efig{Bayes factors for $M$-bin stepwise light curve models with divisible
Dirichlet
priors ($\alpha = 2/M$) vs.\ models with flat priors, for three types of
representative light curve data:  from a flat light curve (red squares),
from a pulse light curve placing all photons in a single phase bin (blue
diamonds),
and from a pulse light curve with a flat background $\approx 9$\% of the
pulse amplitude.
\label{fig:Divis}}

M\"uller raised questions about treatment of two parameters in the
semiparametric pulsar light curve model:  the pulse phase, $\phi$, and the
frequency, $f$.  Rightly noting that the shape prior is shift-invariant,
he asks if $\phi$ may be eliminated altogether.  But the likelihood is
not shift-invariant.  For example, for a particular choice of $M$, there
could be a pulse that is, say, nearly exactly two bins wide.  Depending on
$\phi$, the events from this pulse may be concentrated in two bins, or
spread out over three; the former case has higher likelihood.

Regarding frequency, M\"uller asks, why not ``treat the unknown period as
another parameter.''  This points to a weakness in my description. From a
probabilistic point of view, the frequency is handled as a parameter in the
same manner as other parameters.  The comments at the end of Section~2,
regarding the contrast between Bayesian marginalization over frequency and
frequentist maximization over frequency, extend to how we treat frequency
uncertainty in both the pulsar and exoplanet problems.  It just happens that
there is so much structure in the frequency dimension (nearly as many modes as
Fourier frequencies), that something like exhaustive search is the best way we
currently know of for making sure we find the dominant modes among the dense
forest of modes.  I say ``something like exhaustive search'' because there are
clever ways to explore the frequency parameter without doing the naive search
described before equation~(10).  Atwood et al.\ (2006) show how to use
time-difference tapering to do the search efficiently; Meinshausen et al.\
(2009) describe a more complex but potentially more powerful and general
approach combining tapering with dynamic programming to maximize power subject
to computational resource constraints (and incidentally showing how much may
be gained by having statisticians work on the problem).  In the pulsar
problem, the mode forest is too dense for exploration using standard
Monte Carlo methods. But for the exoplanet problem, Gregory (2007) has
successfully used parallel tempering for frequency search; it requires
millions of likelihood evaluations, indicating it would be unfeasible
for pulsar blind searching (where the number of modes is vastly larger).

For the exoplanet adaptive scheduling problem, M\"uller suggests $m$-step look
ahead procedures may out-perform our myopic procedure, an issue that has
concerned our exoplanet team but which we have yet to significantly explore. The
sequential design folklore that has motivated our efforts to date is that, for
parameter estimation, $m$-step look ahead tends not to yield significant gains
over myopic designs, but that for model comparison, few-step look ahead can
perform significantly better than myopic design.  We have devised a heuristic
few-step look ahead approach for the model comparison problem of planet
detection, but we cannot say yet how much it gains us over myopic designs.  The
earliest expression of the folklore that I have come across is a paper by
Chernoff on sequential design (Chernoff 1961).  He observes: ``The sequential
experimentation problem for estimation\ldots seems to be substantially the same
problem as that of finding `locally' optimal experiments\ldots.  On the other
hand the sequential experimentation problem of testing hypotheses does not
degenerate and is by no means trivial.''  It would be valuable to have more
theoretical insight into the folklore, particularly from a Bayesian perspective.

Finally, M\"uller offers questions and suggestions pertaining to the choice of
utility for orbit estimation and for handling model uncertainty (planet
detection).  Since the observations will ultimately be used by various
investigators for different purposes, some generic measure of information in the
posterior distribution seems appropriate, though with the future use of
inferences being somewhat vague, there cannot be any single ``correct'' choice. 
Our use of Kullback-Leibler divergence (or, equivalently here, Shannon entropy)
is motivated by the same intuition motivating M\"uller's suggestion to 
use precision (which I take to mean inverse variance):  we want the data to tell
us as much about the parameters as possible.  Precision does not appeal to us
because exoplanet posterior distributions can be complex, with significant
skewness, nonlinear correlations, multiple modes, and modes on boundaries of the
parameter space (especially for orbital eccentricity, bounded to $[0,1)$ and
often near a boundary for physical reasons).  In this setting, precision seems
an inadequate summary of uncertainty.  In the limit where the posterior is
unimodal and approximately normal, the entropic measures become the logarithm of
the precision (in the multivariate sense of determinant of the inverse
covariance matrix).  We thus think of these measures as providing a kind of
``generalized precision.''

Noting that the total entropy criterion for the joint estimation/model
comparison problem reduces to separate terms for model and parameter
uncertainty, M\"uller suggests generalizing the criterion to encode
an explicit tradeoff between the estimation and model choice tasks.  This
is an intriguing idea.  At the moment I cannot see obvious astrophysical
criteria that would enable quantification of such a tradeoff.  But
M\"uller's suggestion, along with his observation that sampling cost
is not in our formulation, present me an opportunity to clarify how
complex the actual observing decisions are for astronomers.

Mission
planners for space-based missions, or telescope allocation committees
(TACs) for ground-based observatories, must schedule observations of
{\em many} sources.  For exoplanet campaigns, they will be considering
as-yet unexamined systems, and systems known to have a planet but
with diverse coverage of prior data.  Most exoplanet campaigns share
telescopes with observers pursuing completely different science.
Schedulers must make tradeoffs between science goals within the
exoplanet campaign, and between it and competing science.  There are
costs associated with observations, but there are other nontrivial
constraints as well, such as weather patterns and the phase of the
moon (``dark time'' near the new moon is at a premium; dimmer sources
may be observed then).  In principle one could imagine formal
formulation of the decision problems facing mission planners and TACs,
taking all of these complications into account via utilities or losses.
This may be a worthwhile exercise for a focused mission (e.g., devoted
solely to exoplanet observations); in more general settings the criteria
are probably too hopelessly subjective to allow quantification.  In all
of these settings, we think it would be useful for exoplanet observers
to be able to provide expected information gain versus time calculations,
simply as one useful input for complex scheduling decisions.  M\"uller's
description of our approach as a ``useful default'' is more apt than he may have
realized.  Sequential design is relatively new to astronomy; we hope we can
follow up on some of M\"uller's insightful suggestions as the field moves beyond
these starting points.

\newcommand{\brref}{\section*{References for discussion}
    \begingroup\blef\frenchspacing\absfont}
\newcommand{\erref}{\elef\endgroup}

\brref


\rr
Chernoff, H. (1961). Sequential Experimentation. {\em Bull. Int. Stat.
Inst.} {\bf 38}(4), 3--9.

\rr
Gregory, P. C. (2007). A Bayesian Kepler periodogram detects a second planet in
HD~208487. {\em Mon. Not. Roy. Astron. Soc.} {\bf 374}, 1321--1333.

\erref

\end{document}